\documentclass[acmsmall]{acmart}
\AtBeginDocument{%
  \providecommand\BibTeX{{%
    \normalfont B\kern-0.5em{\scshape i\kern-0.25em b}\kern-0.8em\TeX}}}

\usepackage{multirow}
\usepackage{subcaption}

\newcommand\nnfootnote[1]{%
  \begin{NoHyper}
  \renewcommand\thefootnote{}\footnote{#1}%
  \addtocounter{footnote}{-1}%
  \end{NoHyper}
}

\begin{document}

\title{The Evolution of Distributed Systems for Graph Neural Networks and their Origin in Graph Processing and Deep Learning: A Survey}
\author{Jana Vatter}
\email{jana.vatter@tum.de}
\orcid{0000-0002-5900-5709}
\affiliation{%
  \institution{Technical University of Munich}
  \streetaddress{Department of Computer Science}
  \city{Munich}
  \country{Germany}
}
\author{Ruben Mayer}
\authornote{Parts of the work done while at Technical University of Munich.}
\email{ruben.mayer@uni-bayreuth.de}
\orcid{0000-0001-9870-7466}
\affiliation{%
  \institution{University of Bayreuth}
  \streetaddress{Faculty of Mathematics, Physics \& Computer Science}
  \city{Bayreuth}
  \country{Germany}
}
\author{Hans-Arno Jacobsen}
\email{jacobsen@eecg.toronto.edu}
\orcid{0000-0003-0813-0101}
\affiliation{%
  \institution{University of Toronto}
  \streetaddress{Department of Electrical \& Computer Engineering}
  \city{Toronto}
  \country{Canada}
}

\renewcommand{\shortauthors}{Vatter et al.}

\begin{abstract}
Graph Neural Networks (GNNs) are an emerging research field. This specialized Deep Neural Network (DNN) architecture is capable of processing graph structured data and bridges the gap between graph processing and Deep Learning (DL). As graphs are everywhere, GNNs can be applied to various domains including recommendation systems, computer vision, natural language processing, biology and chemistry. With the rapid growing size of real world graphs, the need for efficient and scalable GNN training solutions has come. Consequently, many works proposing GNN systems have emerged throughout the past few years. However, there is an acute lack of overview, categorization and comparison  of such systems. We aim to fill this gap by summarizing and categorizing important methods and techniques for large-scale GNN solutions. In addition, we establish connections between GNN systems, graph processing systems and DL systems. 
\end{abstract}

\ccsdesc[500]{General and reference~Surveys and overviews}
\ccsdesc[500]{Computing methodologies~Machine learning}
\ccsdesc[300]{Mathematics of computing~Graph algorithms}
\ccsdesc[500]{Computing methodologies~Distributed computing methodologies}
\ccsdesc[500]{Computer systems organization~Neural networks}

\keywords{Graph Neural Networks, Deep Learning Systems, Graph Processing Systems}
\nnfootnote{\copyright \ Owner 2023. This is the author's version of the work. It is posted here for your personal use. Not for redistribution. The definite version was published in ACM Computing Surveys, https://doi.org/10.1145/3597428}

\maketitle

\section{Introduction}

Deep Learning (DL) on graph structured data is a promising and rising field. As graphs are all around us \cite{miller1995wordnet, gupta2013wtf, balaban1985applications}, they can be used in numerous DL applications to model and analyze complex problems. Due to the differing properties of the input data, common DL architectures such as Convolutional Neural Networks (CNNs) \cite{lecun1995convolutional} or Recurrent Neural Networks (RNNs) \cite{rumelhart1986learning, hochreiter1997long} can not easily be used for DL on graphs. Therefore, a new type of Deep Neural Network (DNN) architecture has been developed in the late 2000s, the so-called Graph Neural Network (GNN) \cite{gori2005new, scarselli2008graph}. This architecture bridges the gap between graph processing and DL by combining message passing with neural network (NN) operations. The field of applications of GNNs ranges from recommendation systems \cite{fan2019graph, ying2018graph}, computer vision \cite{gao2020multi, sarlin2020superglue} and natural language processing \cite{yao2019graph, leclair2020improved} to biology and medicine \cite{gaudelet2021utilizing, gao2020mgnn}. 

With the ever-growing size of real world graphs \cite{hu2021ogb} and deeper GNN models \cite{liu2020towards, li2020deepergcn, li2021training}, the need for efficient GNN training solutions has emerged that aim to process large-scale graphs in a fast and efficient manner. With large data sets containing millions of nodes and billions of edges \cite{hu2021ogb}, a high level of parallelization is demanded along with the opportunity to run the computations on distributed architectures. As a result, current research increasingly deals with developing large-scale GNN systems \cite{zheng2020distdgl, wang2021gnnadvisor, jia2020improving}. Multiple issues arise when designing such a system: First, systems originally designed for DL or for graph processing cannot be directly used for GNN training as the former do not support graph processing operations and the latter do not support DL operations.
A vertex in a graph usually only is connected to few other vertices leading to many zero values in the graph adjacency matrix. In contrast, DL operations incorporate high-dimensional features leading to dense matrices. Consequently, both sparse and dense matrix operations need to be supported. Thus, specialized frameworks are being developed incorporating and optimizing both types of tasks. Second, redundant computations and repeated data access might occur during a training iteration. For instance, if nodes share the same neighbor, the neighbors' activation will be computed multiple times in most cases since the computation of the nodes is regarded independently of one another \cite{chen2022regnn, jia2020redundancy}. Thus, the same activation is calculated redundantly. Therefore, the need for appropriate memory management and inter-process communication is enforced. Third, the interdependence of training samples is challenging and leads to increased communication between the machines. One has to decide how to partition and distribute the graph among the machines and which strategy to choose for synchronizing intermediate results.

While there are many works proposing GNN systems that aim to solve the above issues and propose optimization methods, there is an acute lack of overview, categorization, and comparison. There are numerous surveys that classify either scalable graph processing or DL systems, but rarely any articles coping with systems for GNNs. Hence, we aim to fill this gap by giving an overview and categorization of large-scale GNN training techniques.

\subsection{Related Surveys} \label{subsec:related} 
On the one hand, there are surveys on graph processing systems. For instance, Batafari et al. \cite{batarfi2015large} provide an overview of large-scale graph processing systems and present five popular platforms in detail. In addition, they investigate the performance using selected algorithms and graph datasets.
Heidari et al. \cite{heidari2018scalable} and Coimbra et al. \cite{coimbra2021analysis} discuss and categorize graph processing systems in regard to concepts such as partitioning, communication and dynamism.
While the above papers give a general overview of systems irrespective of the hardware set up, Shi et al. \cite{shi2018graph} focus on graph processing on GPUs. They distinguish between graph processing systems on a single GPU and others with a multi GPU setting. 
Another specialized work is done by Kalavri et al. \cite{kalavri2017high} where programming abstractions for large-scale graph processing systems are investigated and evaluated.
Gui et al. \cite{gui2019survey} concentrate on preprocessing, parallel graph computing and runtime scheduling from the hardware side. Xu et al. \cite{xu2014graph} further explore hardware acceleration for graph processing and mainly investigate GPUs.

On the other hand, an overview of systems for DNN training is provided in numerous works. 
Chahal et al. \cite{chahal2020hitchhiker} give an overview of distributed training methods and algorithms. The authors also dive into frameworks for distributed DNN training. 
In their survey, Ben-Nun et al. \cite{ben2019demystifying} give insights into parallelization strategies for DL. They approach the problem from a theoretical angle, model different types of concurrency and explore distributed system architectures.
The survey by Zhang et al. \cite{zhang2018quick} introduces distributed acceleration techniques from the algorithm, distributed system and applications side. In addition, the price and cost of acceleration as well as challenges and limitations are presented. 
Mayer and Jacobsen \cite{mayer2020scalable} investigate challenges, techniques and tools for distributed DL. In their paper, a selection of 11 open-source DL frameworks are analyzed and compared. 
The key aspects of Ouyang et al. \cite{ouyang2021communication} and Tang et al. \cite{tang2020communication} are communication optimization strategies when performing distributed DL. The former highlight algorithm optimizations as well as network-level optimizations for large-scale DNN training, the latter present communication synchronization methods, compression techniques, systems architectures, and different types of parallelization. 
Other research focusing on communication optimizations for large-scale distributed DL is done by Shi et al. \cite{shi2020quantitative}. Instead of examining the scaling problem from a qualitative perspective, the authors pursue a quantitative approach. Therefore, they conduct experiments with selected communication optimization techniques on a GPU cluster. 

To our knowledge, there are many surveys on general aspects of GNNs, like methods, architectures and applications \cite{zhou2020graph, zhang2020deep, hamilton2017representation, chami2020machine, wu2020comprehensive}, but only few study optimizations for GNN training \cite{abadal2021computing, besta2022parallel, serafini2021scalable}. Abadal et al. \cite{abadal2021computing} provide an overview of algorithms and accelerators for GNN training. They focus on methods from the software side as well as hardware specific optimizations, but do not draw parallels to methods introduced by graph processing systems, which is one of our core contributions. Serafini et al. \cite{serafini2021scalable} also explore scalable GNN training. However, they specialize on the comparison of whole-graph and sample-based training and do not investigate other techniques for large-scale GNN training. Besta and Hoefler \cite{besta2022parallel} provide an in-depth analysis of accelerator techniques for GNN training. Their main contribution lies in the field of parallelism techniques whereas we give insight into the overall GNN training pipeline, corresponding optimizations and, most importantly, their origin. In contrast to all of the above, our focus lies on large-scale systems for GNNs and the corresponding optimization techniques within each step of the training process. Additionally, we provide an overview of the two background technologies, systems for graph processing and systems for DNN training, and draw the connection to GNN systems.

\subsection{Our Contributions}
This survey examines distributed systems for scalable GNN training. We differentiate ourselves from other surveys by not only presenting methods for GNN systems, but also by giving insights into the two crucial background topics, systems for graph processing and systems for DNN training. It is important to know about the basic ideas behind those topics to get a better understanding of how and why these techniques are used in large-scale GNN training. We establish connections between the fields and show that most techniques used for large-scale GNN training are inspired by methods either from graph processing or DNN training. This way, we bring together the graph processing, DL and systems community that are all contributing to distributed GNN training. In addition, we categorize and compare the methods regarding questions of partitioning, sampling, inter-process communication, level of parallelism and scheduling. This should make it easier for researchers and practitioners to asses the usability of the presented systems and methods to their own application scenario.

\subsection{Structure of the Survey}
We structure this survey as follows: In Section \ref{sec:back}, we provide fundamentals of distributed systems for graph processing and for DL. We discuss the main aspects for achieving parallelism, scalability and efficiency. In Section \ref{sec:gnn}, we relate the insights from graph processing systems and DL systems to facilitate the comprehension of methods used by systems for GNN training. Finally, open challenges, limitations and research trends are presented in Section \ref{subsec:discussion}.

\section{Foundations} \label{sec:back} 

This section describes relevant topics building the basis of distributed GNN training. We start by giving a short introduction to general graph processing followed by presenting selected graph processing systems. Furthermore, we explore the basic steps of neural network training, and how the training process can be realized in a distributed fashion.

\subsection{Graph Processing}
A graph $G$ is a non-linear data structure, meaning the data elements are not sequentially or linearly ordered. It can be formally described as $G=(V,E)$ where $V$ denotes the set of vertices and $E$ the set of edges. A vertex $v$, also known as node, represents an object and the edge $e$ describes a relation between two vertices. For example, a vertex can represent a person and an edge models the relationship between two persons. There are many different types of graphs, e.g., the graph can contain cycles (cyclic graph) or no cycles (acyclic graph), the edges can be directed or undirected and a weight can be assigned to each edge signifying its importance (weighted graph). Depending on the density of edges within the graph, it can be sparse, dense or fully connected. 

Graphs occur in numerous domains including social networks \cite{tang2010graph, fan2012graph, gupta2013wtf}, route optimization and transportation \cite{sobota2008using, czerwionka2011optimized}, natural language processing \cite{miller1995wordnet, manning2014stanford}, recommendation systems \cite{huang2002graph, silva2010graph, guo2020survey} and the natural sciences \cite{balaban1985applications, ralaivola2005graph, mason2007graph}. For instance, a social network graph models the relation between users, graphs in natural language processing represent the relations between words or sentences, and in the natural sciences, graphs help to model the constitution of molecules.

Graph processing algorithms explore properties of the vertices, relations between them or characteristics of a subgraph or the whole graph. Depending on the use case, different types of graph problems need to be solved by those such as traversals, component identification, community detection, centrality calculation, pattern matching and graph anonymization \cite{dominguez2010discussion, coimbra2021analysis}. Algorithms like depth-first search \cite{tarjan1972depth}, breadth-first search \cite{bundy1984breadth}, Dijkstra \cite{dijkstra1959note}, Girvan-Newman algorithm \cite{girvan2002community}, and PageRank \cite{page1999pagerank} are among the most famous to tackle the presented problems.

\subsection{Distributed Graph Processing} \label{back:distgp}
To process large graphs and to accelerate computation, graph processing is parallelized and distributed across machines. One of the most important methods to process data in parallel is the MapReduce computing model \cite{dean2008mapreduce}. MapReduce incorporates two user-defined primitive functions \texttt{map} and \texttt{reduce}. The \texttt{map} function takes a key-value pair as input and processes it in regard to the specified functionality. All intermediate results are grouped and passed to the \texttt{reduce} function which merges them to obtain a smaller set of values. A master node is responsible for assigning the \texttt{map} and \texttt{reduce} functions to the worker nodes. Each \texttt{map} worker takes a shard of the input pairs, applies the given function and stores the intermediate results locally. The \texttt{reduce} workers load the intermediate files, apply the \texttt{reduce} function and write the results to output files. With the help of the MapReduce computing model, large amounts of data can be handled; however, it reaches its limits when employing it for graph processing. A reason for MapReduce being unsuitable for graph processing is that the vertices within a graph are not independent. When performing a computing step of a graph algorithm, knowledge about multiple vertices is needed, leading to time- and resource-consuming data accesses. While graph processing algorithms may perform multiple iterations, the MapReduce model is optimized for sequential algorithms \cite{mccune2015thinking}. 

Therefore, the Pregel system \cite{malewicz2010pregel} was developed which specializes on parallel graph processing. It supports iterative computations more efficiently than MapReduce and keeps the data set in memory to facilitate repeated accesses. Further, a vertex-centric programming model is provided allowing the user to express graph algorithms more easily. Thus, the user thinks of how the algorithm is modeled from the perspective of a vertex rather than dataflows and transformation operators. During the process, messages are passed between the vertices in a Bulk Synchronous Parallel (BSP) \cite{valiant1990bridging} manner. Hence, the messages are transmitted synchronously after all machines have finished their computation step. The iterative model works as follows: in each iteration, called superstep $S$, a user-defined function is executed in parallel for each vertex $v$. Aside from computing the vertex value, the function can pass messages to other vertices along outgoing edges during an iteration. In superstep $S$, it can send a message which will be received in superstep $S+1$ while messages sent to $v$ in superstep $S-1$ can be read. Similar to MapReduce, Pregel uses a reduction mechanism called "aggregator" to combine resulting values and make them available in the next superstep $S+1$. The introduced synchronous superstep model can be used for various graph algorithms and is the basis for subsequent systems \cite{salihoglu2013gps, khayyat2013mizan, bu2014pregelix}. 

The GAS (\underline{g}ather-\underline{a}pply-\underline{s}catter) model used in PowerGraph \cite{gonzalez2012powergraph} follows the four steps \texttt{gather}, \texttt{sum}, \texttt{apply} and \texttt{scatter}. The \texttt{gather} and \texttt{sum} operations resemble the \texttt{map} and \texttt{reduce} scheme aiming at assembling information about the vertices' neighborhood. This phase is invoked on the edges adjacent to the vertex and executed locally on each machine. The results are sent to the \textit{master} which runs the \texttt{apply} function with the aggregated information. It sends the resulting updates to all machines which execute the \texttt{scatter}-phase. 

Following the two programming models, a general distributed iterative vertex-centric graph processing scheme can be described: Before starting the iterative computation, the graph is split into partitions which are distributed across the machines.
After that, the iterative process starts. Each vertex aggregates messages and applies a given function to compute a new state. Subsequently, the vertex state is updated. Then, the vertex passes messages to adjacent vertices containing the updated information and the aggregation phase starts over.

The vertex-centric paradigm lets the user intuitively define computations on graphs, but it reaches its limitations when processing real-world graphs with skewed degree distributions, large diameters and high densities \cite{yan2014blogel}. Therefore, block-centric models have been developed \cite{yan2014blogel, fan2017grape, tian2013think}. Instead of sending messages from a vertex to its neighbors, blocks comprising of multiple vertices send messages to other blocks. Inside the blocks, information moves freely \cite{tian2013think} resulting in reduced communication cost and improved performance. 

Based on this general iterative model, we present and categorize selected distributed graph processing systems. We start by discussing partitioning strategies and how to store the resulting subgraphs. We distinguish between the synchronization method used in the message propagation phase and show how the messages can be transmitted. Table \ref{tab:GP} summarizes the most important properties of the selected systems and categories.

\begin{table}[t]
\tiny
\caption{Categorization of different distributed graph processing systems}
\label{tab:GP}
\centering
\begin{tabular}{|l|lc|c|c|c|c|c|c|}
    \hline
    & \multicolumn{2}{c|}{} & \multicolumn{2}{c|}{\textbf{Partitioning}} & \multicolumn{2}{c|}{\textbf{Execution Mode}} & \multicolumn{2}{c|}{\textbf{Message Propagation}} \\ \cline{4-5} \cline{6-7} \cline{8-9}
    \textbf{Year} & \textbf{System} & & edge-cut & vertex-cut & synchronous & asynchronous & push & pull \\ \hline \hline 
    
    2010 & \textbf{Pregel} & \cite{malewicz2010pregel} & \checkmark & &  \checkmark & &  \checkmark & \\ \hline 
    
    2012 & \textbf{Apache Giraph} & \cite{apache2012giraph} &  \checkmark & &  \checkmark & &  \checkmark & \\ \hline
    
    2012 & \textbf{GraphLab} & \cite{low2013graphlab, low2014graphlab} &  \checkmark & & &  \checkmark & &  \checkmark \\ \hline 
    
    2012 & \textbf{Distributed GraphLab} & \cite{low2012distributed} &  \checkmark & & &  \checkmark & &  \checkmark \\ \hline 
    
    2012 & \textbf{PowerGraph} & \cite{gonzalez2012powergraph} & &  \checkmark &  \checkmark &  \checkmark & &  \checkmark \\ \hline 
    
    2013 & \textbf{GPS} & \cite{salihoglu2013gps} &  \checkmark & &  \checkmark & &  \checkmark & \\ \hline 
    
    2013 & \textbf{GRACE} & \cite{wang2013asynchronous} &  \checkmark & & &  \checkmark &  \checkmark & \\ \hline 
   
    2015 & \textbf{PowerLyra} & \cite{chen2015powerlyra} &  \checkmark &  \checkmark &  \checkmark &  \checkmark & & \checkmark \\ \hline 

\end{tabular}
\end{table}

\subsubsection{Sampling} \label{back:sampling}
Graph sampling is a preprocessing step aiming to make the graph sparser by removing vertices or edges to reduce processing time and memory consumption. A key challenge here is to ensure that important graph structures are preserved. Iyer et al. \cite{iyer2018bridging} calculate a probability signifying whether an edge between vertex $a$ and vertex $b$ should be kept in the graph or not. The probability value depends on the average degree of the graph, the out-degree of vertex $a$, the in-degree of vertex $b$ and the level of sparsification $s$ which can be chosen by the user. In this manner, less memory is needed while preserving the most important structures within the graph. 
The fast, approximate ASAP engine \cite{iyer2018asap} for graph pattern mining consists of two phases, namely the \textit{sampling} and \textit{closing} phase. In the \textit{sampling} phase, the graph is treated as a stream of edges. The edges are either randomly selected or depending on the previously streamed ones. Then, the \texttt{closing} phase awaits certain edges to complete patterns. This technique helps to preserve certain structures within the graph while excluding certain edges.  
An application-aware approach that drops certain messages is proposed by Schramm et al. \cite{schramm2022flexible}. In every superstep, a given percentage of messages is identified as least important and dropped on-the-fly. The calculation of the importance value is dependent on the desired application and the deployed graph algorithm. Therefore, this method can be used in a plethora of applications.

\subsubsection{Partitioning} \label{back:part}
There exist numerous approaches on graph partitioning algorithms that the systems adopt and refine. Those algorithms either follow the edge-cut \cite{karypis1998fast, stanton2012streaming, tsourakakis2014fennel}, vertex-cut \cite{schlag2019scalable, hanai2019distributed, petroni2015hdrf, mayer2018adwise, mayer2021hybrid, mayer2022out} or hybrid-cut model \cite{fan2022application, chen2015powerlyra}. Further, we distinguish between offline and online partitioners. While offline partitioners \cite{karypis1997metis, hendrickson1993chaco} determine the partitions according to the whole graph, online partitioners \cite{tsourakakis2014fennel, petroni2015hdrf, mayer2018adwise} stream vertices or edges and assign the vertex or edge to a partition on-the-fly. An interesting approach is HEP \cite{mayer2021hybrid} where offline and online components are combined into a hybrid scheme. 

To asses the quality of the partitions, metrics like \textit{replication factor}, \textit{communication cost} and \textit{workload balancing} are used. The \textit{replication factor} indicates the ratio between the number of replicas and the total number of vertices. Based on that, the \textit{communication cost} can be determined. Whenever an edge is cut, communication between the partitions is needed. When using an edge-cut method for example, the communication increases proportional to the number of edges that are cut. The so-called \textit{workload balancing} aims to partition the graph in a way such that during the computation phase of the graph processing algorithm the load is balanced among the workers.

There exist various cost functions that help to form partitions. Depending on the goal and the application, the cost function needs to be defined differently. In the Linear Deterministic Greedy (LDG) \cite{stanton2012streaming} method, a vertex is allocated to the partition where it has most edges. This is combined with a penalty function indicating the capacity of a partition. Consequently, communication costs are minimized and balanced workloads across partitions is ensured. 
FENNEL \cite{tsourakakis2014fennel} unifies two heuristics, a vertex is either assigned to the cluster that shares the largest number of neighbors or the one with the smallest number of non-neighbors. This results in a minimal number of edges that are cut and consequently minimal communication costs. 
The High Degree (are) Replicated First (HDRF) \cite{petroni2015hdrf} method handles graphs with skewed distribution where the degree of the nodes highly varies. The goal is to balance the load evenly by cutting and replicating high-degree vertices. 
2PS \cite{mayer20202ps} gathers clustering information in a preprocessing phase which is then used in the scoring mechanism. 
Instead of defining one particular cost function, Fan et al. \cite{fan2022application} use an application-driven approach. A cost model based on a given application algorithm is minimized in order to partition the graph. This results in an adaptive partitiong strategy suitable for numerous use cases.

\begin{figure}[t]
    \centering
    \begin{subfigure}[r]{0.17\textwidth}
        \includegraphics[width=\textwidth]{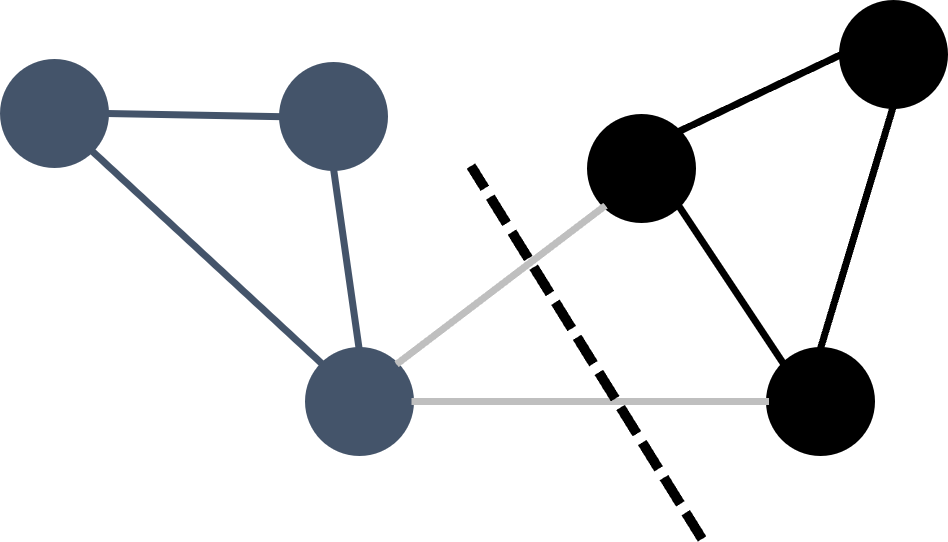}
        \caption{edge-cut}
        \label{fig:edge}
    \end{subfigure}
    \hspace{30pt}
    \begin{subfigure}[l]{0.17\textwidth}
        \includegraphics[width=\textwidth]{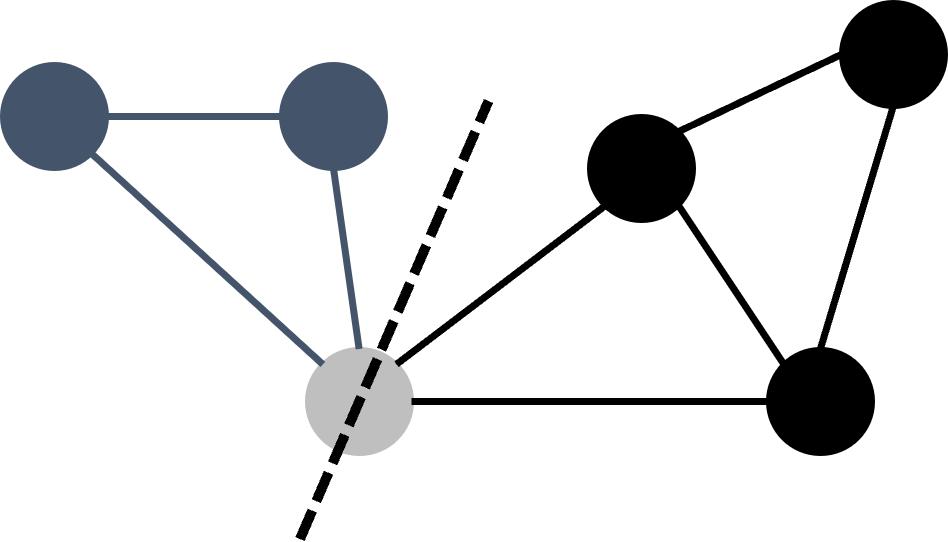}
        \caption{vertex-cut}
        \label{fig:vertex}
    \end{subfigure}
    \caption{Edge-cut and vertex-cut partitioning.}
    \label{fig:edgevertexcut}
\end{figure} 

Pregel \cite{malewicz2010pregel} uses one of the simplest methods to initially partition the graph. Here, the partitioning is based on the vertex ID using a hash function $hash(ID)\,mod\,N$ where $N$ is the number of partitions. The partitions are then evenly distributed across all workers. The underlying partitioning principle is called edge-cut as the edges are cut and replicated while the vertices are assigned to the different machines (Fig. \ref{fig:edge}). Other systems relying on edge-cut partitioning are Apache Giraph \cite{apache2012giraph}, GraphLab \cite{low2013graphlab, low2014graphlab}, Distributed GraphLab \cite{low2012distributed}, Graph Processing System (GPS) \cite{salihoglu2013gps} and GRACE \cite{wang2013asynchronous}. Compared to other partitioning strategies, the presented edge-cut methods induce less computation cost and less overheads. Conversely, the following methods aim to reduce the cut size at the cost of higher runtimes and memory overheads. However, the obtained partitions are of improved quality.

With edge-cut partitioning, the number of vertices is balanced across partitions, however, the number of edges per partition might highly vary. To obtain more equal partitions and to improve distributed processing of natural graphs, Gonzalez et al. propose balanced vertex-cut partitioning (Fig. \ref{fig:vertex}) in their PowerGraph \cite{gonzalez2012powergraph} model. Natural graphs typically are imbalanced and thus, difficult to partition. Some vertices have a high degree of outgoing edges while other vertices are of low degree. Hence, the computation costs per partition vary a lot. For that reason, the system employs balanced vertex-cut partitioning. The novel partitioning strategy improves the processing of natural graphs by equally distributing the edges between the machines whereas adjacent vertices are mirrored on the machines. One vertex among all replicas is randomly assigned as the \textit{master} to compute and update the vertex state. All \textit{mirrors} keep local copies of the vertex state. 

A dynamic repartitioning strategy to reduce communication is introduced in GPS \cite{salihoglu2013gps}. First, the system partitions the input graph with a standard partitioning technique. During the execution of the graph processing algorithm, communication is observed to determine which vertices to reassign to another worker and when to do so. Messages that are passed over the network are significantly reduced because repeated vertex accesses are performed by workers that already loaded the vertex in a previous iteration. Therefore, processing time is sped up, the workload is balanced and scaling to larger graphs is improved. 

PowerLyra \cite{chen2015powerlyra} further optimizes the processing of natural graphs by introducing a hybrid-cut partitioning algorithm which combines edge- and vertex-cut. Furthermore, it handles high-degree vertices and low-degree vertices separately to minimize the replication of edges and vertices. The proposed balanced p-way hybrid-cut algorithm exploits a form of edge-cut if a vertex has a low number of outgoing edges and vertex-cut if a vertex is of high degree. 

GridGraph \cite{zhu2015gridgraph} introduces a grid representation for graphs to speed up partitioning. First, the vertices are partitioned into $P$ chunks where each chunk contains connected vertices. The edges are sorted into the resulting $P \times P$ grid according to their source and destination vertices. In this method, the list of edges does not need to be ordered leading to small preprocessing times. Further, the grid representation can facilitate the execution of the following graph processing steps.

As there are numerous partitioning strategies with different characteristics and objectives, it is challenging to choose the optimal one for a given application. Therefore, some experimental studies investigate the performance and resource usage of different strategies \cite{gill2018study, pacaci2019experimental, abbas2018streaming} while the EASE system \cite{merkel2023partitioner} provides a quantitative prediction and automatic selection.

\subsubsection{Inter-Process Communication} \label{back:inter}
Once the graph has been partitioned, it has to be decided how the worker processes synchronize their data, e.g., by sending messages or by accessing shared memory. In the former case, the systems store the subgraphs locally on the assigned machines and exchange synchronization messages. GraphLab \cite{low2013graphlab, low2014graphlab} supports the latter case and uses a shared-memory abstraction. A data graph accessible for all workers stores the program state as well as the corresponding data structures. Oriented on GraphLab, PowerGraph \cite{gonzalez2012powergraph} also performs computation following a shared-memory view.

\subsubsection{State Synchronization and Scheduling} \label{back:syncgp}
When iteratively executing graph processing algorithms, the computation steps can be performed following a synchronous, asynchronous or hybrid scheme. The underlying basic principle Pregel follows is the BSP model. Every node goes through an iteration by aggregating and combining the desired features and subsequently updating the state. When the node has finished the computation step, it waits for all other nodes to finish before continuing with the next iteration. This assures that every node shares the same parameters. 

While the synchronous mode works well in a lot of cases, it can be inefficient in other cases. An example is the label propagation algorithm for community detection \cite{raghavan2007near}. Here, each vertex is assigned an initial label. In each iteration, the vertex adopts the label the maximum of neighbors has. After some propagation rounds through the network, dense communities consent to the same label. If the synchronous mode is chosen and the graph is bipartite, meaning each vertex of one subgraph connects to each vertex in another subgraph, the labels might oscillate and change after each iteration. This makes it impossible for the algorithm to terminate, as the labels therefore need to be stable. To solve this issue, asynchronous execution is used. Asynchronous processing means that vertices can already read state updates of the current iteration in addition to the updates of the previous iterations. In GNN training, the asynchronous mode has the ability to prioritize specific state updates which results in faster convergence of the overall computation. Another important benefit is the avoidance of long idle times through stragglers.
 
GraphLab \cite{low2013graphlab, low2014graphlab} uses an asynchronous execution method. Three steps are performed independently by each worker: \texttt{fold}, \texttt{merge} and \texttt{apply}. In the \texttt{fold} step, the data across all vertices is gathered. If provided, a \texttt{merge} operation is performed. Otherwise, the \texttt{apply} function directly completes the computation and stores the data in shared memory. This process is done without regarding the phase of computation the other workers are currently passing. GRACE \cite{wang2013asynchronous} allows for high performance execution by adapting the BSP model to permit asynchronous processing. Hence, idle times are minimized, but it is possible that the workers perform their iteration with stale data. 
As both synchronization methods have their benefits and drawbacks, PowerGraph \cite{gonzalez2012powergraph} allows the user to either choose a synchronous or asynchronous mode. The synchronous mode is performed analogously to the synchronous execution in Pregel, the asynchronous mode resembles the GraphLab computation model.

Usually, choosing a mode manually requires the user to deeply understand the graph processing system and often does not lead to the optimal performance. Hysync \cite{xie2015sync} removes this issue by automatically switching between the synchronous and asynchronous mode according to a set of heuristics. The heuristics aim to predict the performance of the current mode and determine the step at which a switch to the other mode can be beneficial. 
In the Adaptive Asynchronous Parallel (AAP) model \cite{fan2020adaptive}, each worker decides on its own when to start the next computation step depending on two parameters. The first parameter is the relative progress of a worker compared to the other ones. The second parameter indicates data staleness. Consequently, stragglers are avoided while also reducing stale computations. 

PowerLyra \cite{chen2015powerlyra} goes a step further and not only provides both synchronization modes, but also distinguishes between high- and low-degree vertices to determine how they are processed. The former are processed based on the GAS model (i.e. \texttt{gather}, \texttt{apply}, \texttt{scatter}) \cite{gonzalez2012powergraph}. 
The master vertex activates the mirrors to execute the \texttt{gather} function and the results are sent back to the master. After having received all messages, the master runs the \texttt{apply} function. A combined message with the updated data and the activation for the \texttt{scatter} function is sent from master to mirror. In contrast to the original GAS model, PowerLyra combines the \texttt{apply} and \texttt{scatter} messages from master to mirror vertices to minimize communication. The system handles the low-degree vertices similar to GraphLab. The master vertex performs the \texttt{gather} and \texttt{apply} phase locally. Hereafter, activation and update messages are combined and sent to the mirrored vertices. Each mirror then performs the \texttt{scatter} phase. Because of the adapted scheme, replication of edges is eliminated and in each iteration, only one message needs to be sent by a mirror.

Opposed to the synchronous mode, using an asynchronous execution implies the need of a scheduling scheme which can influence the convergence of the overall computation. Besides, a high level of concurrent execution can be achieved as scheduling helps to determine the order of the tasks and assigns the tasks to the machines. GraphLab \cite{low2013graphlab, low2014graphlab} provides the so-called \textit{set scheduler}. Based on the dependencies between the tasks, an \textit{execution plan} is established and an overall speed-up of the computations can be observed. 
In their GRACE \cite{wang2013asynchronous} system, Wang et al. incorporate a customizable scheduling policy. The system loosens the restrictions of the BSP model and lets the user prioritize specific vertices for faster convergence. One can choose an individual set of vertices and also the desired processing order of those. Then, the system calculates a \textit{scheduling priority} for each vertex to determine the overall execution order. 

\subsubsection{Message Propagation} \label{back:message}
Regardless of the execution mode and scheduling technique, messages need to be transmitted between the vertices containing the updated states. There are two main methods to transfer messages: \textit{push} and \textit{pull}. While Pregel-based systems \cite{malewicz2010pregel, apache2012giraph, salihoglu2013gps, wang2013asynchronous} use the \textit{push}-operation, systems supporting the GAS model \cite{low2013graphlab, low2014graphlab, low2012distributed, gonzalez2012powergraph, chen2015powerlyra} rely on \textit{pulling} the messages. After each iteration, Pregel-based systems synchronously propagate the update messages. All vertices simultaneously \textit{push} their messages, meaning each vertex directly sends a message to all its adjacent vertices containing the updated state. 
GraphLab \cite{low2013graphlab, low2014graphlab}, on the other hand, stores the data graph with associated features in shared memory. Like this, all workers are able to access the data any time. At the beginning of an iteration, the worker \textit{pulls} the current data graph via the \texttt{gather} operation to perform calculations on the most recent features. At the end of an iteration, the worker updates the data graph with the newly computed state.

\subsection{Distributed Neural Network Training} \label{back:nntraining}
A neural network consists of a number of connected nodes, called neurons, organized in one or multiple layers. Each neuron takes an input and processes it in regard to given weights and an update function. When having computed the new value, an activation function \cite{sharma2017activation, ramachandran2017searching} decides how important the output value is. The values are passed through the network until the last set of neurons is reached, called forward pass. After that, a loss function is computed in regard to the calculated and expected output. In order to minimize the loss, the weights need to be adapted. The backpropagation \cite{rumelhart1995backpropagation} algorithm is used to adjust weights in a backward pass through the network. A widely used technique to do so is stochastic gradient descent (SGD) \cite{robbins1951stochastic, ruder2016overview}. It computes the partial gradients by considering the calculated loss. With the help of those gradients, the weights are adjusted. This is called backward pass. The whole process, forward and backward pass, is iteratively repeated until convergence or a maximum number of iterations is reached. The final neural network weights can be used to make predictions on unseen data \cite{widrow199030, lawrence1993introduction, goodfellow2016deep}. 

With the growing amount of training data and the increase of model size, distributed neural network training has become necessary. In the following, we discuss different techniques to scale neural network training with regard to parallelism, execution mode and coordination. Table \ref{tab:NNOverview} gives an overview of the described systems and techniques.

\begin{table}[b]
\tiny
\caption{Categorization of systems for distributed neural network training}
\label{tab:NNOverview}
\centering
\begin{tabular}{|l|lc|c|c|c|c|c|c|}
    \hline
    
    & \multicolumn{2}{c|}{} & \multicolumn{2}{c|}{\textbf{Parallelism}} & \multicolumn{2}{c|}{\textbf{Synchronization Mode}} & \multicolumn{2}{c|}{\textbf{Coordination}} \\ \cline{4-5} \cline{6-7} \cline{8-9}
    
    \textbf{Year} & \textbf{System} & & data &  model & synchronous & asynchronous  & centralized & decentralized \\ \hline \hline
    
    2012 & \textbf{DistBelief} & \cite{dean2012large} & \checkmark & & & \checkmark & \checkmark & \\ \hline
    
    2014 & \textbf{Project Adam} & \cite{chilimbi2014project} & & \checkmark & & \checkmark & \checkmark & \\ \hline
    
    2015 & \textbf{MALT} & \cite{li2015malt} & \checkmark & & &  \checkmark & & \checkmark \\ \hline
    
    2016 & \textbf{Tensorflow} & \cite{abadi2016tensorflow} & \checkmark & \centering \checkmark & \checkmark & \checkmark & \checkmark & \\ \hline 
    
    2016 & \textbf{Ako} & \cite{watcharapichat2016ako} & \checkmark & & & \checkmark & & \checkmark \\ \hline
    
    2019 & \textbf{CROSSBOW} & \cite{koliousis2019crossbow} & \checkmark & & \checkmark & & & \checkmark \\ \hline

    2019 & \textbf{PyTorch} & \cite{paszke2019pytorch} & \checkmark & \checkmark & \checkmark & \checkmark & \checkmark & \checkmark \\ \hline 

\end{tabular}
\end{table}

\subsubsection{Parallelism} \label{back:par}
Data parallelism allows for parallel training on multiple processors. Therefore, the data is divided into a fixed number of subsets. Each worker is assigned a subset which it processes on a local replica of the model. After that, the resulting model parameters are exchanged with the other workers. The next iteration is performed with the updated parameter configuration. This process is repeated until convergence.
Systems using data parallelism are for instance MALT \cite{li2015malt} and Ako \cite{watcharapichat2016ako}. 
If the model itself is too big to fit on a single machine \cite{krizhevsky2012imagenet, brown2020language}, it is split up. In model parallelism, the computation nodes process the whole data set on their partition of the model. After having finished the computation, the intermediate output of the forward pass is passed to the machines responsible for computing the subsequent layer. Here, scheduling is important to efficiently coordinate the training process. The most intuitive way to partition the model is layer-wise, meaning each layer is assigned to one node. However, this sometimes does not benefit parallelism as the worker controlling the current layer needs to wait for the worker handling the prior layer to finish before being able to run their computation. A system exploiting model parallelism is Project Adam \cite{chilimbi2014project} where each machine is assigned a certain part of the model. Another possibility to distribute the model more intelligently is according to its specific architecture. Suitable could be architectures like the two-fold Siamese network \cite{he2018twofold} where some of the components can be easily run in parallel. 

The two techniques are combined, e.g., by PipeDream \cite{narayanan2019pipedream} and GPipe \cite{huang2019gpipe}, into the so-called hybrid or pipeline parallelism. Here, the data as well as the model are shared among the workers. In contrast to PipeDream, the GPipe algorithm further splits the input mini-batches into chunks to maximize the number of concurrent calculations within an iteration. In general, the use of hybrid parallelism can significantly improve training speed in comparison to data and model parallelism.

\subsubsection{Synchronization Mode} \label{back:syncdl}
When training neural networks in a distributed fashion, it is important that the machines regularly exchange parameter updates or intermediate results to ensure convergence. This can either follow a synchronous, asynchronous or hybrid mode \cite{dean2012large, chen2016revisiting, jin2016scale}. When using a synchronous mode, the updates are sent simultaneously to the other nodes as soon as all machines have finished their computation. Then, the nodes continue with the computation of the next iteration. In this manner, every machine is always aware of the current parameters. A downside of this approach is that a single straggler decreases the speed of the whole training process \cite{cipar2013solving}. A system following the synchronous execution mode is CROSSBOW \cite{koliousis2019crossbow}. 

Models using asynchronous execution \cite{dean2012large, chilimbi2014project, li2015malt, watcharapichat2016ako} eliminate the problem of stragglers by not waiting for all workers to finish. Instead, the updates are sent as soon as they are available. Thus, the training speed is increased and the resources are efficiently used. A drawback of this method is that the worker nodes might not always be up to date, therefore computing their updates on stale parameters. The computation of gradients on outdated parameters can lead to a slower or no convergence at all which is called stale gradients problem \cite{dutta2018slow}. 
An attempt to merge the synchronous and asynchronous execution scheme was made by Ho et al. \cite{ho2013more}. Their model is based on synchronous execution, but incorporates a staleness threshold determining how many time steps two workers may be apart until the faster worker needs to wait for the slower worker to finish its current computation. In contrast to synchronous execution, this decreases the impact of stragglers while also limiting the staleness of the parameters to ensure up-to-date computation. 
Popular frameworks like TensorFlow \cite{abadi2016tensorflow} and PyTorch \cite{paszke2019pytorch} leave the choice of synchronization level to the user. Here, both methods are supported.

\subsubsection{Coordination} \label{back:coord}
Besides determining the synchronization mode, it needs to be decided how the parameters are stored and the updates are coordinated. Common ways to do so are in a centralized or decentralized manner. The centralized method operates with a global parameter server which stores, aggregates and distributes the relevant updates. Consequently, all machines share the same set of parameters. However, the use of a parameter server introduces a single-point of failure as it is part of all update requests. DistBelief \cite{dean2012large} and Project Adam \cite{chilimbi2014project} are systems operating in a centralized fashion. 
Decentralized systems \cite{li2015malt, watcharapichat2016ako, koliousis2019crossbow} eliminate the need for a parameter server by passing the update messages directly from machine to machine by using a collective communication primitive such as an \texttt{all-reduce} operation \cite{sanders2019sequential}. Here, each machine exchanges update information with its peers and combines the received parameters with its own. As a result, each machine holds the latest set of parameters. An advantage of this approach is that the computation of updates is balanced among all machines \cite{koliousis2019crossbow}. As both strategies have their benefits and drawbacks, PyTorch \cite{paszke2019pytorch} leaves the choice of coordination to the user.

To conclude, there are several ways to perform distributed DNN training. Depending on the architecture and the data, one might choose between data, model or hybrid parallelism, synchronous or asynchronous updates and a centralized or decentralized system.
In all methods, parameters need to be exchanged between the machines.
Therefore, communication is a bottleneck that needs to be addressed when training neural networks in a multi-machine setting. In addition, it is important that the resources are fully utilized without long idle times \cite{zhang2020network}.

\section{Systems for Graph Neural Networks} \label{sec:gnn}

\subsection{Graph Neural Network Basics}
The term graph neural network (GNN) initially emerged in a work by Gori et al. \cite{gori2005new} and was further investigated by Scarselli et al. \cite{scarselli2008graph}. It refers to neural network architectures that do not take multiple independent or sequenced data samples as input like CNNs \cite{lecun1995convolutional} or RNNs \cite{sherstinsky2020fundamentals}, but rather graphs. In contrast to images or text, graphs do not follow a specific structure and are not sequentially ordered. A graph $G$ can be formally denoted as $G= (V,E)$. It consists of a set of vertices $V$ and a set of edges $E$. A vertex $v$ represents an object and is also known as node. An edge $e$ describes the relation between two vertices. As graphs are unstructured, it is necessary to employ a new type of neural network, the GNN. They combine DNN operations like matrix multiplication and convolution with methods known from graph processing like iterative message propagation \cite{jia2020improving, wang2021flexgraph}. Due to uniting DNN operations and message passing, GNNs are sometimes also denoted as Message Passing Neural Networks \cite{gilmer2017neural, riba2018learning, zhang2020conv, hamilton2020graph}. 

To begin with, each vertex of an input graph is initially represented by a feature vector, called activation. This initial activation only incorporates information about the vertex itself but not about the context within the graph. At each layer, a set of DNN operations and message passing steps are performed vertex-wise to update the activation values. In the first step of an iteration, each vertex aggregates the activations of its adjacent vertices by exchanging messages according to an \textit{aggregation} function
\begin{equation}
    a_v^{(k)} = AGGREGATE^{(k)}(\{h_u^{(k-1)}|u \in N(v)\})
\end{equation} where $a_v^{(k)}$ denotes the aggregated values of vertex $v$ at the $k$-th layer. The term $h_u^{(k-1)}|u \in N(v)$ describes the activations of the neighboring vertices at the previous layer with $N(v)$ denoting the neighbors of $v$ in the given graph. Next, the gathered information is combined and the current value of the vertex is updated with an \textit{update} function. The \textit{update} function $h_v{(k)}$ can include standard DNN operations like matrix multiplication and is defined by
\begin{equation}
    h_v^{(k)} = UPDATE^{(k)}(a_v^{(k)}, h_v^{(k-1)})
\end{equation}
where $h_v^{(k)}$ is the new activation of vertex $v$ at the $k$-th layer. To obtain the updated activation, the aggregated activations $a_v^{(k)}$ are combined with the vertices' activation of the previous layer $h_v^{(k-1)}$. A special case occurs if $k=1$, then the initial activations $h_v{(0)}$ or $h_u{(0)}|u \in N(V)$ are needed, respectively \cite{jia2020improving, hamilton2020graph}. The new values now serve as starting point for the next layer where the activations are aggregated and combined again. This process is repeated iteratively. Consequently, more and more vertices are explored. After $k$ layers, the $k$-hop neighborhood of a vertex is captured. When the final layer has been passed, the final representation of a vertex includes information about the vertex itself as well as about the vertices within the graph context. 

\begin{figure}[t]
    \centering
    \includegraphics[width=0.8\columnwidth]{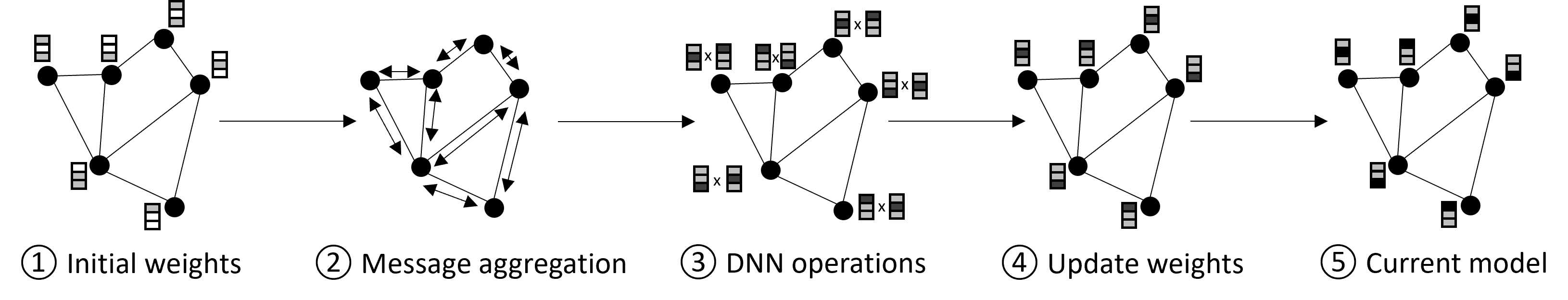}
    \caption{Schematic of the GNN training process}
    \label{fig:gnntraining}
\end{figure}

In Figure \ref{fig:gnntraining}, an overview of a complete forward pass is given. It consists of the above described steps: (1) fetch the initial weights, (2) pass and aggregate messages from neighboring nodes, (3) perform DNN operations and (4) update the weights according to a given function. The steps (2) to (4) are repeatedly executed. After $n$ iterations, a final model configuration (5) is obtained. 

Analogous to the general neural network training described in Section \ref{back:nntraining}, a loss function is computed relative to the output of the forward pass. Backpropagation is applied during the backward pass in order to adapt the weights of the network \cite{xu2018powerful}. 
After multiple forward and backward passes, the model can make vertex- and edge-level predictions. In order to make assertions about the whole graph, a pooling layer needs to be added which aggregates and combines all states and labels contained in the output graph based on a specified pooling operation \cite{zhou2020graph}. 

There exist several types of GNNs \cite{wu2020comprehensive}, among the most famous ones are Gated Graph Neural Networks (GG-NN) \cite{li2015gated}, Graph Convolution Networks (GCN) \cite{kipf2016semi}, GraphSAGE \cite{hamilton2017representation}, Graph Attention Networks (GAT) \cite{velivckovic2017graph} and Graph Auto-Encoders (GAE) \cite{kipf2016variational}. Architectural differences involve the message propagation process, sampling method, pooling operation as well as the composition of the layers. For instance, the GraphSAGE model uses a max-pooling strategy while GCNs use mean-pooling instead. GATs include masked self-attention in the pooling process and GG-NNs capture spatial and temporal changes throughout time by using gated recurrent units as update module. In general, if the relations between objects are essential to make predictions on the data, GNNs can be used and are often preferable over common DNN architectures \cite{lecun1995convolutional, hochreiter1997long}. In contrast to DNNs which read in the data object by object or in an ordered sequence, GNNs naturally capture the relations within a graph and are also able to predict the relations between data points. This cannot be easily done with other DNN models. Thus, GNNs benefit the processing of and prediction on graph data. Sometimes it can be advantageous to combine GNNs with other DNN models, for instance, when handling temporal graphs where nodes and edges are updated sequentially \cite{kumar2019predicting, rossi2020temporal, zhang2021cope}. Here, a GNN is combined with a Recurrent Neural Network. However, if there are no significant connections between the individual data points, there usually is no need to choose a GNN over another DNN model to perform DL.

\subsection{Categorization of Methods for Distributed Graph Neural Network Training} \label{gnn:methods}
Training GNNs on large graphs is a challenging task which requires high communication, large memory capacity and high bandwidth \cite{md2021distgnn}. In contrast to general distributed DNN training, all data points within a graph are connected and not independent of each other. Due to this dependency, one can not simply split the data to process the batches in parallel. Furthermore, memory-intensive edge-centric operations in addition to arithmetic-intensive vertex-centric operations need to be regarded when optimizing GNNs \cite{kiningham2020grip}. Thus, large-scale graph processing methods or efficient DNN training operations can not directly be used for GNN training. More specialized techniques adjusted to GNN characteristics are needed to overcome the described challenges and to speed up training and inference. 

\begin{figure} [t]
    \centering
    \includegraphics[width=0.7\columnwidth]{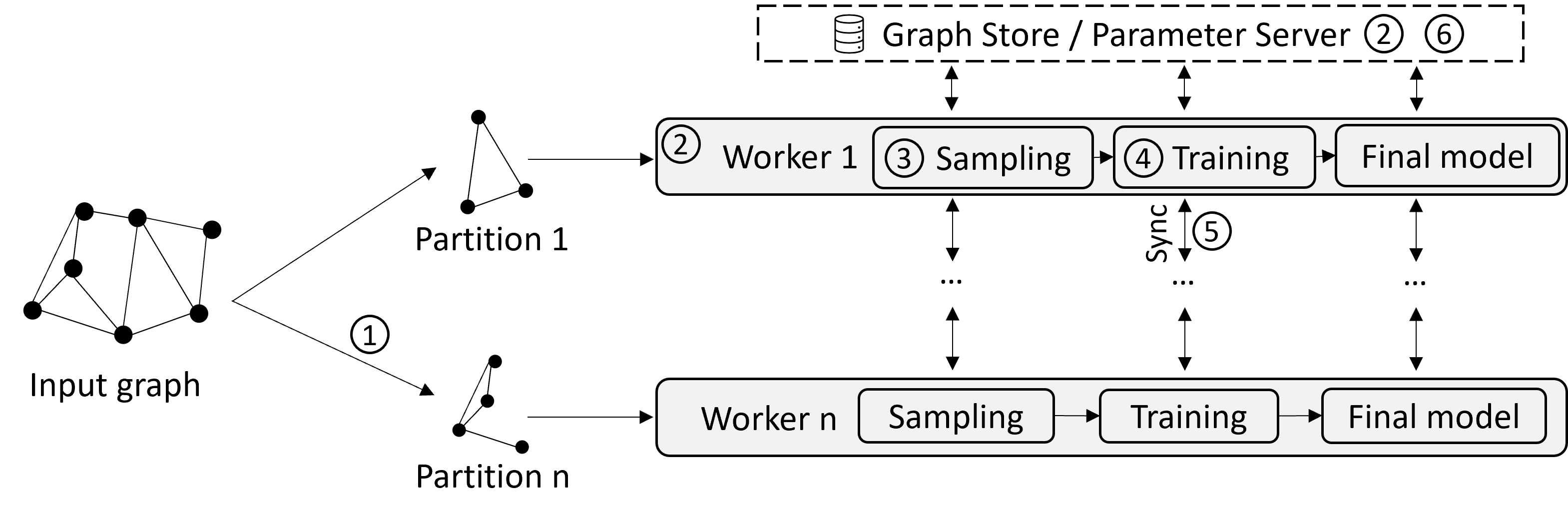}
    \caption{General set up of a GNN system}
    \label{fig:systemsoverview}
\end{figure}

A general set up of GNN systems is shown in Figure \ref{fig:systemsoverview}. Similar to the first step in distributed graph processing, the graph can initially be partitioned and distributed across the workers (\textcircled{1}: Section \ref{subsubsec:part}). Other types of parallelism (Section \ref{subsubsec:parallel}) are also possible, however data parallelism is the most common choice. After that, questions of how to store the subgraphs and whether to cache any data need to be addressed (\textcircled{2}: Section \ref{subsubsec:inter}). This can either be done locally on the machines or globally on a dedicated graph store. Depending on the training mode, a sampling step is performed (\textcircled{3}: Section \ref{subsubsec:sampling}). Here, only a selection of vertices instead of the whole partition is used to train the model. The main training begins with a random initial representation of the vertices. Then, messages are pulled to or pushed from adjacent vertices (\textcircled{4}: Section \ref{subsubsec:message}), DNN operations are applied and the updated vertex states are shared with the other vertices before the next iteration starts. The whole process can be executed in a synchronous or asynchronous fashion (\textcircled{5}: Section \ref{subsubsec:sync}) and various scheduling techniques may be applied (Section \ref{subsubsec:schedule}). Instead of a decentral \texttt{all-reduce} operation to share the parameters, a centralized parameter server might be used (\textcircled{6}: Section \ref{subsubsec:coord}). After having finished all iterations, predictions can be made based on the final set of parameters. Most systems additionally provide a programming abstraction adapted to the individual optimizations (Section \ref{subsubsec:api}). In the following, we will present and categorize systems for GNNs based on the described steps for setting up such a system.

\subsubsection{Partitioning} \label{subsubsec:part} 
Graph processing systems rely on partitioning the input graph and distributing it across machines before starting the main computation. GNN systems adopt the idea of partitioning because the input graphs are unlikely to fit in a single machine's memory. Some systems use traditional edge-cut or vertex-cut methods \cite{zheng2020distdgl, md2021distgnn} whereas others combine those with features like a cost model \cite{jia2020improving}, feasibility score \cite{lin2020pagraph} or dataflow partitioning \cite{kiningham2020greta}. Table \ref{tab:partitioning} summarizes the different partitioning methods. 

\begin{table}[b]
\tiny
\caption{Categorization of partitioning strategies}
\label{tab:partitioning}
\centering
\begin{tabular}{|l|lc|c|c|c|c|c|c|c|p{0.25\columnwidth}|}
    \hline
    & \multicolumn{2}{c|}{} & \multicolumn{3}{c|}{\textbf{Cut type}} & \multicolumn{2}{c|}{\textbf{Static vs. Dynamic}} & \multicolumn{2}{c|}{\textbf{Offline vs. Online}} & \\ \cline{4-6} \cline{7-8} \cline {9-10}
    
    \textbf{Year} & \textbf{System} & & \centering Edge & \centering Vertex & \centering Hybrid & \centering Static & \centering Dynamic & \centering Offline & \centering Online & \textbf{Balancing Objective} \\ \hline \hline
    
    2019 & \textbf{NeuGraph} & \cite{ma2019neugraph} & & & \centering \checkmark & \centering \checkmark & & \centering \checkmark & & 2D partitioning with equally-sized disjoint vertex chunks \\ \hline 

    2019 & \textbf{GReTA} & \cite{kiningham2020greta} & & & \centering \checkmark & \centering \checkmark & & \centering \checkmark & & 2D dataflow partitioning \\ \hline 
    
    2020 & \textbf{ROC} & \cite{jia2020improving} & \centering \checkmark & & & & \centering \checkmark & \centering \checkmark & & runtime of a partition \\ \hline 
    
    2020 & \textbf{AGL} & \cite{zhang2020agl} & \centering \checkmark & & & \centering \checkmark & & \centering \checkmark & & neighborhood size \\ \hline 
    
    2020 & \textbf{PaGraph} & \cite{lin2020pagraph} & \centering \checkmark & & & & \centering \checkmark & & \centering \checkmark & feasibility score, node degree, computation expense \\ \hline
    
    2020 & \textbf{DistDGL} & \cite{zheng2020distdgl} & \centering \checkmark & & & \centering \checkmark & & \centering \checkmark & & minimum edge-cut \\ \hline 
    
    2021 & \textbf{P\textsuperscript{3}} & \cite{gandhi2021p3} & \centering \checkmark & & & \centering \checkmark & & \centering \checkmark & & random hash \\ \hline
    
    2021 & \textbf{GNNAdvisor} & \cite{wang2021gnnadvisor} & \centering \checkmark & & & \centering \checkmark & & \centering \checkmark & & 2D workload partitioning \\ \hline 
    
    2021 & \textbf{DistGNN} & \cite{md2021distgnn} & & \centering \checkmark & & \centering \checkmark & & & \centering \checkmark& edges per partition \\ \hline 
    
    2021 & \textbf{DeepGalois} & \cite{hoang1050efficient} & \centering \checkmark & & & & & & \centering \checkmark & user-defined policy \\ \hline 
    
    2021 & \textbf{ZIPPER} & \cite{zhang2021zipper} & & \centering \checkmark & & \centering \checkmark & & \centering \checkmark & & 2D partitioning with equally-sized disjoint vertex chunks \\ \hline 
    
\end{tabular}
\end{table}

In DistDGL \cite{zheng2020distdgl}, the input graph is partitioned using the METIS \cite{karypis1997metis, karypis1998fast} edge-cut algorithm. In addition, optimizations like multi-constraint partitioning \cite{karypis1998multilevel} and a refinement phase are used to improve load balancing. DistGNN \cite{md2021distgnn} generates partitions with a minimum vertex-cut algorithm and the tool Libra \cite{xie2014distributed}. Here, edges belong to one specific partition while vertices can correspond to multiple partitions, hence they need to be replicated. The so-called \textit{replication factor} measures the number of replicas. A lower \textit{replication factor} induces less communication across partitions. Moreover, Libra generates balanced partitions by equally distributing edges among subgraphs. 

NeuGraph \cite{ma2019neugraph} first preprocesses the input graph with the METIS edge-cut algorithm and then applies a grid-based partitioning scheme which combines edge- and vertex-cut. It is similar to the method used in GridGraph \cite{zhu2015gridgraph} and assigns each vertex and its corresponding features to one of $P$ vertex chunks. Then, the adjacency matrix is tiled into $P \times P$ chunks containing the corresponding edges. This partitioning strategy benefits the edge-wise processing, because only the source and destination vertex data need to be loaded. 
Unlike NeuGraph, GReTA \cite{kiningham2020greta} does not partition the graph itself, but the dataflow into blocks. The dataflow, also called nodeflow, is a graph structure representing the propagation of feature vectors throughout the forward pass of the GNN model. The vertices each represent a mathematical unit of computation while the edges represent the inputs and outputs of the units. At first, GReTA partitions the vertices of the dataflow graph into $n$- and $m$-sized chunks. Then, blocks of size $n \times m$ are formed out of the adjacency matrix containing the relevant edges. Hence, only a part of the grid representation instead of the whole graph needs to be loaded for performing a computation step. Additionally, an entire column can be processed in the aggregation phase. Inspired by NeuGraph and GReTA, ZIPPER \cite{zhang2021zipper} also uses a grid-based partitioning technique where the graph adjacency matrix is tiled into multiple rectangular blocks. It is distinguished between source and destination vertices throughout the partitioning process to ensure that each block is only associated with one source and one destination partition. Thus, each edge is uniquely identified. As described by the former systems, this partitioning strategy is applied to reduce the memory footprint and communication. 

Usually, the graph is partitioned at the beginning of the computation and these partitions are used until the end of the overall process. ROC \cite{jia2020improving}, however, repartitions the graph before each iteration using an online regression model. A cost model predicts the execution time of various operations on a given graph partition based on parameters like the number of vertices and edges and the number of memory accesses. The cost model is updated and minimized at the end of each iteration with the actual runtimes needed for the subgraph. Then, the graph is repartitioned based on the updated costs. 

Zhang et al. provide AGL (Ant Graph ML system) \cite{zhang2020agl} to employ large GNNs for industrial use cases. The \textit{GraphFlat} module is responsible for dividing the input graph into $k$-hop neighborhoods. It uses a pipeline inspired by message passing to produce the desired neighborhoods. In a MapReduce-style, self-information about a vertex is generated, propagated along outgoing edges and aggregated. This is done until $k$ iterations are reached. Now, the nodes contain information determining the partitions regarding their $k$-hop neighbors. 

PaGraph \cite{lin2020pagraph} designs a GNN-aware partitioning algorithm which distributes the vertices among partitions depending on a score. In each iteration of the algorithm, a vertex is scanned and a vector is calculated where the $i$-th position determines the feasibility for assigning the vertex to partition $i$. The score incorporates features of the vertices so far assigned to partition $i$, the in-neighbor set of the vertex and the expected number of vertices in the partition. The current vertex is allocated to the partition with the highest feasibility score and it is proceeded with the subsequent vertex. Balanced partitions and an evenly distributed computing effort are ensured. 
GNNAdvisor \cite{wang2021gnnadvisor} relies on neighbor partitioning where only the adjacent vertices of a target vertex belong to the given partition. A reason to choose neighbor partitioning over edge- or vertex-cut partitioning is the mitigation of highly varying partition sizes. Further, the probability for tiny partitions is lower which reduces the managing costs. 
In contrast to the above presented systems, DeepGalois \cite{hoang1050efficient} allows for customized partitioning by incorporating the Customizable Streaming Partitioner (CuSP) \cite{hoang2019cusp} framework. A simple API lets the user determine the specific partitioning policy, supported are edge-cut, vertex-cut and hybrid-cut. Hence, the user can tailor the partitioning strategy to the specific GNN architecture. 
Unlike systems exploiting a compute-intensive customized partitioning strategy \cite{kiningham2020greta, jia2020improving, lin2020pagraph}, P\textsuperscript{3} \cite{gandhi2021p3} relies on a simple random hash partitioner. This ensures simple, fast and efficient partitioning with only minimal overhead. Here, the initial graph partitioning helps to balance the workload, but the main optimizations to scale to large graphs are done in the consecutive steps of the system, namely Sampling (Section \ref{subsubsec:sampling}), Inter-Process Communication (Section \ref{subsubsec:inter}), choice of Parallelism (Section \ref{subsubsec:parallel}), Synchronization Mode (Section \ref{subsubsec:sync}), Scheduling (Section \ref{subsubsec:schedule}) and Coordination (Section \ref{subsubsec:coord}).

\subsubsection{Sampling} \label{subsubsec:sampling} 
The underlying idea of sampling originates in graph processing. However, the idea of sampling slightly differs in connection with GNN training. Different to DNN training, the samples within a graph are not independent. When performing mini-batch training, one can not randomly pick out a number of vertices without regarding the relation to other ones. Thus, training samples of a mini-batch need to include the k-hop neighborhood of a vertex. However, without sampling, these neighborhoods are likely to "explode" as the neighborhood size quickly grows with each hop. For that reason, numerous strategies such as vertex-wise \cite{hamilton2017inductive}, layer-wise \cite{zou2019layer} or subgraph-based \cite{chiang2019cluster} sampling are introduced. The underlying idea of all these methods is to restrict the number of k-hop neighbors to be explored to prevent the described neighborhood explosion issue \cite{zheng2020distdgl}. While sampling is an established technique working well for many tasks, the choice of the specific strategy depends on the desired downstream ML task, the graph structure and the objective of the sampling method \cite{rozemberczki2020little}. 

An early algorithm for efficient sampling in GNN training is GraphSAGE (SAmple and aggreGatE) \cite{hamilton2017inductive}. The model trains mini-batches and restricts the neighborhood size per layer. The number of vertices to be sampled is fixed and the vertices are randomly chosen. GraphSAGE is able to work on larger graphs compared to the general GCN architecture. However, by picking the vertices at random, neighborhood information might be lost leading to a decrease in accuracy.  
Therefore, PinSage \cite{ying2018graph} adopts importance-based neighborhood sampling. The new technique incorporates random walks to compute a score for each vertex and select a fixed number of neighbors accordingly. Hence, memory usage can be controlled and adjusted if needed, while yielding higher accuracy than random sampling.

FastGCN \cite{chen2018fastgcn} further explores the idea of sampling based on a calculated score. The authors introduce an importance based layer-wise sampling mechanism where an importance score and a fixed neighborhood size determine which vertices to select. The score mainly depends on the degree of each vertex and is calculated for each layer to restrict the corresponding number of vertices. Thus, the neighborhood explosion problem can be avoided and large graphs can be handled. However, as the vertex-wise importance is calculated independently per layer, the selected neighborhoods for two subsequent layers may differ which might lead to slow convergence. 
This issue is faced by LADIES \cite{zou2019layer} which exploits importance sampling in a layer-dependent way. Depending on the sampled vertices in the previous layer, the neighboring vertices in the current layer are selected and a bipartite graph between the two layers is constructed. After that, the sampling probability is calculated and a fixed number of vertices is sampled. This procedure is repeated for each layer to sample the needed vertices. 

ClusterGCN \cite{chiang2019cluster} allows for subgraph-based sampling. Contrary to general mini-batch GCN training where the vertices are randomly chosen, a graph clustering algorithm is used to form the mini-batches. The clustering strategy aims at minimizing the number of links between vertices in the batch or between multiple batches. As a consequence, ClusterGCN is faster and uses less memory compared to previous methods. 
Zeng et al. propose GraphSAINT \cite{zeng2019graphsaint} which also supports subgraph-based sampling. In contrast to former sampling-based systems first building a GCN and then sampling the input graph, GraphSAINT starts with sampling subgraphs and builds a GCN for each subgraph after that. By building a complete GCN for each sample, extensions like skip connections as proposed by JK-Net \cite{xu2018representation} are applicable without needing to adapt the sampling process. JK-Net requires the samples of the current layer to be a subset of the previous layers' samples which is naturally met by using GraphSAINT. Further, GraphSAINT ensures a minimized neighborhood size while maintaining a high accuracy. 

AliGraph \cite{yang2019aligraph} incorporates sampling by providing three steps: \texttt{traverse}, \texttt{neighborhood} and \texttt{negative}. \texttt{Traverse} draws a set of vertices and edges from the subgraphs and \texttt{neighborhood} is responsible for building a vertex context which may be one- or multi-hop neighbor vertices. The last step, \texttt{negative}, speeds up convergence of the training by setting up negative samples. Here, negative sampling refers to including exemplary vertices in the training process that are not part of the given sample. For instance, given a Graph $G$ where vertex $A$ is connected to vertex $B$, but there is no edge between vertex $A$ and vertex $C$. Negative sampling would mean to not only incorporate edge $(A,B)$ in the training process as positive example, but also edge $(A,C)$ with an explicit negative tag indicating there is no connection between those vertices. 

Inspired by ClusterGCN \cite{chiang2019cluster}, GraphTheta \cite{li2021graphtheta} uses a clustering algorithm to form the graph samples. Within one subgraph, the algorithm detects and builds communities maximizing intra-community and minimizing inter-community edges. Due to forming communities before the main sampling step, the sampled vertices tend to overlap not as much as with (random) neighbor sampling leading to fewer repeated vertex accesses. Despite the advantage of less redundant calculations, graphs with weak community structures are not supported and the batch size can be imbalanced due to varying community sizes. 

The AGL system \cite{zhang2020agl} provides a variety of sampling methods that can be chosen from, for instance, uniform sampling and weighted sampling. This ensures that the user can select the best strategy for each application. Wang et al. also pursue the idea of providing several sampling techniques in DGL \cite{wang2019deep}. A set of methods is provided, like the well-known neighbor sampling \cite{hamilton2017inductive} and cluster sampling \cite{chiang2019cluster}. The DistDGL system \cite{zheng2020distdgl} is based on DGL and supports several sampling techniques, but implements the sampling step in a distributed way. The sampling request originates from the trainer process and is sent to the machine responsible for the target set of vertices. After having received the request, the sampling worker calls the sampling operators of DGL and performs sampling on the local partition. It sends the results back to the trainer process which generates a mini-batch by assembling all acquired results. 
Like AGL, DGL and DistDGL, P\textsuperscript{3} \cite{gandhi2021p3} does not provide a particular sampling strategy. However, it adopts the sampling method given by the GNN architecture. If no specific method is included, P\textsuperscript{3} proceeds without a sampling phase. This ensures that a variety of GNN architectures are supported by the system. 

Although sampling-based methods may decrease training time of GNNs, there remain issues like lack of consistency \cite{hu2020open} and limited applicability to GNN architectures with many-hop or global context layers. For that reason, NeuGraph \cite{ma2019neugraph} and ROC \cite{jia2020improving} omit the sampling phase and rely on full-batch training. A short overview of the different sampling strategies can be found in Table \ref{tab:sampling}.

\begin{table}[t]
\tiny
\caption{Categorization of sampling strategies}
\label{tab:sampling}
\centering
\begin{tabular}{|l|lc|c|c|c|c|c|p{0.25\columnwidth}|}
    \hline
    & \multicolumn{2}{c|}{} & \multicolumn{3}{c|}{\textbf{Method}} & \multicolumn{2}{c|}{\textbf{Coordination}} & \\ \cline{4-6} \cline{7-8}

    \textbf{Year} & \textbf{System} & & \centering Community & \centering User Defined & \centering Full Batch & \centering Centralized & \centering Distributed & \textbf{Main Sampling Concepts} \\ \hline \hline

    2019 & \textbf{DGL} & \cite{wang2019deep} & & \centering \checkmark & & \centering \checkmark & & choose method based on application \\ \hline 

    2019 & \textbf{NeuGraph} & \cite{ma2019neugraph} & & & \centering \checkmark & \centering \checkmark & & no sampling \\ \hline 

    2019 & \textbf{Aligraph} & \cite{yang2019aligraph} & &\centering \checkmark  & & \centering \checkmark & & three steps: traverse, neighborhood, negative \\ \hline 

    2020 & \textbf{ROC} & \cite{jia2020improving} & & & \centering \checkmark & \centering \checkmark & & no sampling \\ \hline 

    2020 & \textbf{AGL} & \cite{zhang2020agl} & & \centering \checkmark & & \centering \checkmark & & choose method based on application \\ \hline 

    2020 & \textbf{DistDGL} & \cite{zheng2020distdgl} & & \centering \checkmark & & & \centering \checkmark & sampling worker responsible for local partition \\ \hline 

    2021 & \textbf{GraphTheta} & \cite{li2021graphtheta} & \centering \checkmark & & & \centering \checkmark & & sample from clusters, minimize overlapping vertices \\ \hline 

    2021 & \textbf{P\textsuperscript{3}} & \cite{gandhi2021p3} & & \centering \checkmark & & \centering \checkmark & & adapt to given GNN architecture and application \\ \hline 

\end{tabular}
\end{table}

\subsubsection{Programming Abstractions} \label{subsubsec:api} 
To facilitate the implementation of the desired GNN architecture and to support custom optimizations, the proposed systems come with programming abstractions including user-defined functions. There are programming models based on message passing \cite{wang2019deep, fey2019fast, hu2020featgraph}  while other abstractions are using a dataflow paradigm \cite{ma2019neugraph, kiningham2020greta, li2021graphtheta}. 

A PyTorch \cite{paszke2019pytorch} extension tailored to GNN training is PyTorch Geometric (PyG) \cite{fey2019fast}. The library comes with a message passing base class where the user only needs to define the construction of messages, the update function as well as the aggregation scheme. Message propagation is handled automatically. Numerous GNN architectures can be implemented, for instance, GCN \cite{kipf2016semi}, SGC \cite{wu2019simplifying}, GraphSAGE \cite{hamilton2017inductive}, GAT \cite{velivckovic2017graph}, and GIN \cite{xu2018powerful}. 
The Deep Graph Library (DGL) \cite{wang2019deep} also lets the user define the desired GNN model as a set of message passing primitives covering forward and backward pass. The central abstraction is the graph data structure \textit{DGLGraph}. (Pre-)defined functions such as neighbor sampling directly operate on a \textit{DGLGraph} and return a subgraph object. Therefore, manually slicing tensors and manipulating graph data is made obsolete in contrast to frameworks such as PyG. 
Another message passing-based programming interface is introduced by FeatGraph \cite{hu2020featgraph}. In addition to customizing the GNN model, the user is able to determine the parallelization strategy for the vertex- and edge-wise feature dimension computation. 

For applying the optimization techniques proposed by P\textsuperscript{3} \cite{gandhi2021p3}, the system provides a message passing API that developers can use to implement GNN models that automatically include the optimizations. The \texttt{P-TAGS} API consists of six functions: \texttt{partition}, \texttt{scatter}, \texttt{gather}, \texttt{transform}, \texttt{sync} and \texttt{apply}. All functions are user-defined and target a different step in the GNN training process. 
An independent \texttt{partition} function is provided where the developer can implement an individual partitioning algorithm to reduce communication. While \texttt{scatter} is a message passing function defined on each edge, \texttt{gather} assembles the messages vertex-wise with a commutative and associative function. The \texttt{transform} function applies the given NN operations on each vertex to compute the partial activations. Then, a neighborhood representation is computed with NN operations such as convolution. Those representations are collected over the network with \texttt{sync} and a composite \texttt{apply} function is used to update the vertices' states. 

Oriented on the GAS model \cite{gonzalez2012powergraph}, NeuGraph \cite{ma2019neugraph} introduces the SAGA-NN (scatter-applyedge-gather-applyvertex with Neural Networks) programming scheme. The vertex-centric model expresses one GNN forward pass as four steps: \texttt{scatter}, \texttt{applyedge}, \texttt{gather}, and \texttt{applyvertex}. \texttt{scatter} and \texttt{gather} are predefined and responsible for data propagation while the other two, \texttt{applyedge} and \texttt{applyvertex}, are defined by the user and are expressed as dataflow with tensor-based operations. In the first phase, namely \texttt{scatter}, information about the vertices is passed onto adjacent edges where the values are aggregated and subsequently combined to form a single edge value in \texttt{applyedge}. The \texttt{gather} step propagates the updated values to the vertices where they are assembled. The vertex state is updated in the last step, \texttt{applyvertex}. Within each step, the computation is parallelized. The abstraction combines graph processing and NN training by uniting the dataflow model with a vertex-centric view. In general, SAGA-NN follows the common iterative GNN computation model which makes it applicable to various architectures \cite{kipf2016semi, sukhbaatar2016learning, li2015gated}.

To allow for efficient execution of GNN training on an accelerator, GReTA \cite{kiningham2020greta} introduces four stateless user-defined functions: \texttt{\underline{g}ather}, \texttt{\underline{re}duce}, \texttt{\underline{t}ransform} and \texttt{\underline{a}ctivate}. \texttt{gather} loads and aggregates edge and vertex data. The \texttt{reduce} operation merges the data into one single value. The current and previous reduced vertex state are combined in \texttt{transform}. Finally, \texttt{activate} updates the vertices with new values. A GNN layer is expressed as one or multiple GReTA programs making it expressive enough for a diverse set of GNN models \cite{kipf2016semi, bresson2017residual, hamilton2017inductive, xu2018powerful}. 
\texttt{NN-TGAR} proposed by GraphTheta \cite{li2021graphtheta} provides user-friendly programming and enables training on a cluster of machines. Moreover, it unites graph processing and graph learning frameworks. In contrast to the GAS model and GReTA, GraphTheta lets the user implement the GNN model in a vertex- and edge-centric view. While some functions are applied to each vertex, other functions are applied to each edge. The first step of the abstraction, the \texttt{NN-T} operation, transforms values vertex-wise and generates corresponding messages. After that, \texttt{NN-G} (\texttt{gather}) is applied to each edge to update the edge values and transmit the message to the destination vertex. The \texttt{sum} operation in turn operates on each vertex and combines the received messages with methods like averaging or concatenation (\texttt{NN-A}). Then, the result is applied to the vertices and the parameters are updated with \texttt{reduce}. \texttt{NN-T}, \texttt{NN-G} and \texttt{NN-A} are implemented as neural networks, making it easy to perform the forward pass and subsequent gradient computation. In addition, an encoding layer is decomposed into subsequent independent stages allowing for general applicability. 

FlexGraph \cite{wang2021flexgraph} introduces the programming model NAU (\texttt{neighborselection}, \texttt{aggregation} and \texttt{update}). In contrast to models based on the GAS abstraction \cite{gonzalez2012powergraph, ma2019neugraph}, NAU comprises \texttt{neighborselection} which builds Hierarchical Dependency Graphs (HDGs) including chosen neighbors to capture the dependencies among vertices. After that, the neighborhood features are aggregated and a neighborhood representation is computed in the \texttt{aggregation} step. In the \texttt{update} phase, a new representation is calculated consisting of old features and the new neighborhood representation. Moreover, one single message comprises of multiple features and messages are assembled to reduce traffic. As opposed to programming models like SAGA-NN \cite{ma2019neugraph} and its variants, NAU is not limited to 1-hop neighbors during computation. Additional to flat aggregation operations, hierarchical aggregation can be used to support various GNN architectures. Therefore, NAU also supports GNN models with indirect neighbors and hierarchical aggregations, for instance, PinSage \cite{ying2018graph}, MAGNN \cite{fu2020magnn}, P-GNN \cite{you2019position}, JK-Net \cite{xu2018representation}, while SAGA-NN merely supports architectures where direct neighbors and flat aggregations are regarded \cite{kipf2016semi, xu2018powerful, marcheggiani2017encoding}.  

\begin{table}[b]
\tiny
\caption{Categorization of programming abstractions}
\label{tab:progabstr}
\centering
\begin{tabular}{|c|cc|p{0.22\columnwidth}|p{0.22\columnwidth}|p{0.22\columnwidth}|}
    \hline
    
    \textbf{Year} & \textbf{System} & & \textbf{Expressiveness} & \textbf{Optimizations} & \textbf{Algorithms} \\ \hline \hline
    
    2019 & \textbf{DGL} & \cite{wang2019deep} & message passing abstraction & operates on DGLGraph & GCN, SGC, GraphSAGE, GAT, GIN \\ \hline
    
    2019 & \textbf{PyG} & \cite{fey2019fast} & message passing abstraction & optimized sparse softmax kernels & GCN, SGC, GraphSAGE, GAT, GIN, ARMA, APPNP \\ \hline 
    
    2019 & \textbf{GReTA} & \cite{kiningham2020greta} & dataflow abstraction & one or multiple GReTA programs per GNN layer & GCN, G-GCN, GraphSAGE, GIN \\ \hline
    
    2019 & \textbf{NeuGraph} & \cite{ma2019neugraph} & dataflow model with vertex-centric view, direct neighbors and flat aggregations & tensor-based operations & CommNet, GCN, GG-NN \\ \hline 
    
    2020 & \textbf{FeatGraph} & \cite{hu2020featgraph} & message passing abstraction & custom parallelization strategy & GCN, GraphSAGE, GAT \\ \hline
    
    2021 & \textbf{GraphTheta} & \cite{li2021graphtheta} & vertex- and edge-centric abstraction & independent steps are implemented as neural networks & GCN, FastGCN, VR-GCN \\ \hline 
    
    2021 & \textbf{FlexGraph} & \cite{wang2021flexgraph} & indirect neighbors and hierarchical aggregations & hierarchical dependency graphs & PinSage, MAGNN, P-GNN, JK-Net \\ \hline 
    
    2021 & \textbf{P\textsuperscript{3}} & \cite{gandhi2021p3} & vertex- and edge-centric abstraction & user-defined partition function & S-GCN, GCN, GraphSAGE, GAT \\ \hline 
    
    2021 & \textbf{Seastar} & \cite{wu2021seastar} & vertex-centric abstraction & improved usability & GCN, GAT, APPNP, R-GCN \\ \hline 
    
\end{tabular}
\end{table}

Inspired by Pregel \cite{malewicz2010pregel}, the underlying programming model of Seastar \cite{wu2021seastar} is realized in a vertex-centric fashion. From the viewpoint of a vertex, the user defines functions to implement the GNN architecture. Seastar then executes the given operations on each vertex. This improves usability compared to message passing systems \cite{fey2019fast, wang2019deep, ma2019neugraph} and dataflow programming systems \cite{ma2019neugraph, yang2019aligraph, alibaba2020euler}. With the proposed abstraction, it is possible to implement GNN models more easily and the implementation can be adjusted faster. More details on the various programming abstractions are shown in Table \ref{tab:progabstr}.

\subsubsection{Inter-Process Communication} \label{subsubsec:inter} 
Before starting the GNN training process, how to store the partitions, samples and corresponding features and whether to cache any vertex data needs to be determined. DGL \cite{wang2019deep} is built on top of DNN frameworks like TensorFlow \cite{abadi2016tensorflow} or PyTorch \cite{paszke2019pytorch} and leaves the memory management those frameworks instead of developing its own storing and caching strategy. Even though those frameworks are able to store datasets used for DNN training efficiently, more specialized techniques are beneficial for GNN training. Here, samples might contain overlapping neighborhoods, some vertices are repeatedly accessed, and it is crucial to preserve the connections within the graph. Therefore, the following systems employ more sophisticated methods which are summarized in Table \ref{tab:interprocess}.

\begin{table}[b]
\tiny
\caption{Categorization of inter-process communication methods}
\label{tab:interprocess}
\centering
\begin{tabular}{|l|lc|c|c|p{0.15\columnwidth}|p{0.15\columnwidth}|p{0.15\columnwidth}|}
    \hline
    & \multicolumn{2}{c|}{} & \multicolumn{2}{c|}{\textbf{Storage}} & \multicolumn{2}{c|}{\textbf{Caching}} & \\ \cline{4-7}
    
    \textbf{Year} & \multicolumn{2}{c|}{\textbf{System}} & \centering Centralized & \centering Distributed & \centering Data & \centering Objective & \textbf{Optimizations} \\ \hline \hline
    
    2019 & \textbf{DGL} & \cite{wang2019deep} & \centering \checkmark & \centering \checkmark & - & - & leaves memory management and caching to base framework \\ \hline 
    
    2019 & \textbf{Aligraph} & \cite{yang2019aligraph} & & \centering \checkmark & neighbors of selected vertices & based on importance value & -  \\ \hline 
    
    2020 & \textbf{ROC} & \cite{jia2020improving} & & \centering \checkmark  & intermediate tensors on GPU & minimize cost model based on the graph, GNN model and GPU device & - \\ \hline 
    
    2020 & \textbf{PaGraph} & \cite{lin2020pagraph} & \centering \checkmark & & frequently accessed feature vectors & minimize computation and communication & - \\ \hline 
    
    2020 & \textbf{DistDGL} & \cite{zheng2020distdgl} & & \centering \checkmark & - & - & KVStore, co-location of data and computation \\ \hline 
    
    2021 & \textbf{GraphTheta} & \cite{li2021graphtheta} & & \centering \checkmark & - & - & task-oriented layout \\ \hline 
    
    2021 & \textbf{P\textsuperscript{3}} & \cite{gandhi2021p3} & \centering \checkmark & & graph and/or features & user-defined & store partial activations \\ \hline 
    
    2021 & \textbf{GNNAdvisor} & \cite{wang2021gnnadvisor} & & \centering \checkmark & - & - & vertex reordering \\ \hline 
    
\end{tabular}
\end{table}

AliGraph \cite{yang2019aligraph} proposes to cache neighbors of important vertices. An \textit{importance} value for each vertex with respect to the number of k-hop in- and out-neighbors is calculated. The out-neighbors of vertices with an \textit{importance} value exceeding a user-defined threshold are locally cached. In this way, frequently needed data becomes easily accessible and communication cost is reduced. 
ROC \cite{jia2020improving} optimizes runtime performance by caching intermediate tensors on GPUs while maintaining the remaining data in the host memory. By caching these tensors, the data transfers between CPU and GPU are decreased. A cost model is built with respect to a given input graph, a GNN model and a GPU device. This cost model is minimized with a dynamic programming algorithm to find the globally optimal caching strategy. 
As the data copy operation from CPU to GPU is a major bottleneck in distributed GNN training, PaGraph \cite{lin2020pagraph} uses a computation-aware caching mechanism to minimize data copy. Vertex features as well as structural information about the graph are stored in a \textit{Graph Store Server} on the CPU. This shared memory is globally accessible. Additionally, a cache on each GPU keeps frequently accessed feature vectors. Before deciding which vertices to cache, Lin et al. analyze the characteristics of the training process. GNNs including a sampling technique randomly shuffle the samples in each epoch making it impossible to predict at runtime which vertices belong to which mini-batch. As a consequence, it is not possible to forecast which vertices are accessed at the next training iteration. However, the out-degree of a vertex indicates how likely it is to be sampled in the whole epoch. With a higher out-degree, it is an in-vertex for a higher number of neighbors. Thus, those vertices are chosen more often in other samples, yield higher computation costs and should be easily accessible. The caching policy of PaGraph is oriented on those findings. It pre-sorts the vertices by out-degree and fills up the cache following that order. This leads to a high cache hit ratio and reduced data transfer. 

A distributed file system is used by AGL \cite{zhang2020agl} to store the neighborhoods. During computation, one or a batch of them is loaded instead of the whole graph. This highly decreases communication between graph stores and workers. AGL can be run on a single machine or a CPU cluster. DistDGL \cite{zheng2020distdgl} also works on multiple CPUs. Thus, the graph structure, corresponding features and embeddings are stored on multiple machines. The so-called distributed key-value store (\textit{KVStore}) is globally accessible for all trainer processes. DistDGL co-locates data and computation, meaning the distribution of vertices and edges among the \textit{KVStore} servers on the machines resemble the obtained graph partitions in the partitioning step. Consequently, trainer processes can directly access the data and communication is reduced. Euler \cite{alibaba2020euler} also exploits a distributed storage architecture. The graph engine layer is responsible for loading and dividing the graph into subgraphs and distributing them among the machines. 

A more specialized memory optimization is introduced by GNNAdvisor \cite{wang2021gnnadvisor}. The underlying idea is to couple vertices and the computing units where they are processed more tightly. Therefore, graph reordering is performed as by RabbitOrder \cite{arai2016rabbit} and the \textit{GO-PQ} algorithm \cite{wei2016speedup}. Neighbor groups being close to each other are assigned consecutive vertex IDs increasing the possibility for them to be scheduled closely on the same machine. 
As two adjacent neighborhoods often share common neighbors, the L1 cache is more efficiently used and data locality is exploited. ZIPPER \cite{zhang2021zipper} also includes a vertex reordering technique. Here, a heuristic degree sorting strategy is used to group the out-edges of the source vertices. As a consequence, vertex data is more efficiently reused and redundancies are minimized. 

GraphTheta \cite{li2021graphtheta} also stores subgraphs in a distributed fashion. To achieve low-latency access and reduced memory overhead, the proposed parallel tensor storage utilizes a task-oriented layout. The memory needed for each task, e.g., forward pass, backward pass or aggregation phase, is grouped process-wise. The task-specific memory includes raw data and tensors which are further sliced into frames for more efficient access. For each frame, memory is allocated and deallocated immediately after usage throughout the whole computation process to reduce the memory utilization. 

Gandhi et al. \cite{gandhi2021p3} extend the KVStore introduced in DistDGL \cite{zheng2020distdgl}. In addition to vertex and edge data, P\textsuperscript{3} also stores partial activations in the KVStore. The extended KVStore coordinates data movement across machines. As soon as the machines are synchronized, the accumulated activation is moved to the device memory and shared with the trainer process. Additionally, P\textsuperscript{3} allows the user to define a caching strategy. A simple method tested by the authors is to store the input on a minimum number of machines and replicate partitions on so far unused machines.

\subsubsection{Parallelism} \label{subsubsec:parallel}
In distributed DNN training, the appropriate type of parallelism can scale computations to large sets of data. The proposed methods, namely data parallelism, model parallelism and hybrid parallelism, have proven to work well. For that reason, researchers apply methods taken from DNN training to GNN training. 

The most common method is data parallelism. Analog to parallelism in DNN training, the graph, is split into subgraphs. The model is replicated on the machines while each machine handles its own subgraph. Systems like NeuGraph \cite{ma2019neugraph}, PaGraph \cite{lin2020pagraph} and DistGNN \cite{md2021distgnn} rely on data parallelism. 
Two different implementations of data parallelism, namely edge and vertex parallelism, are introduced by DGL \cite{wang2019deep}. Two types of matrix multiplications are distinguished to determine whether to process the data edge- or vertex-parallel. Vertex-parallel computation is used for generalized sparse-dense matrix multiplication. In this case, the entire adjacency list of a vertex is managed by one thread. The edge-parallel strategy is used for generalized sampled dense-dense matrix multiplication where one edge is managed by one thread. Here, the workload is balanced implicitly while the workload extent of vertex-parallel processing depends on the vertex degree.  

A hybrid between data and model parallelism is employed in P\textsuperscript{3} by Gandhi et al. \cite{gandhi2021p3} to tackle issues like ineffectiveness of partitioning and GPU underutilization. First, the model is partitioned and distributed among the machines. After having computed the partial activations for Layer 1, the machines apply a \texttt{reduce} function to aggregate those activations. Then, P\textsuperscript{3} switches to data parallelism to finish the forward pass. The backward pass is very similar, until Layer 1, data parallelism is exploited and the error gradient is exchanged among the machines. Then, P\textsuperscript{3} switches back to model-parallel execution to perform the remaining steps of the backward pass locally.

\begin{table}[t]
\tiny
\caption{Categorization of types of parallelism}
\label{tab:parallel}
\centering
\begin{tabular}{|l|l|l|c|c|c|l|}
    \hline
    
    & \multicolumn{2}{c|}{} & \multicolumn{3}{c|}{\textbf{Type of Parallelism}} & \\ \cline{4-6}
    
    \textbf{Year} & \multicolumn{2}{c|}{\textbf{System}} & \centering Data & \centering Model & \centering Hybrid & \textbf{Main Concepts} \\ \hline \hline
    
    2019 & \textbf{DGL} & \cite{wang2019deep} & \centering \checkmark & & & Edge and vertex parallelism \\ \hline
    
    2019 & \textbf{NeuGraph} & \cite{ma2019neugraph} & \centering \checkmark & & & Mini-batch training  \\ \hline 
    
    2020 & \textbf{PaGraph} & \cite{lin2020pagraph} & \centering \checkmark & & & Mini-batch training \\ \hline
    
    2021 & \textbf{GraphTheta} & \cite{li2021graphtheta} & & & \centering \checkmark & Data and operations are split \\ \hline 
    
    2021 & \textbf{P\textsuperscript{3}} & \cite{gandhi2021p3} & & & \centering \checkmark & Push-pull parallelism \\ \hline
    
    2021 & \textbf{DistGNN} & \cite{md2021distgnn} & \centering \checkmark & & & Mini-batch training \\ \hline
    
    2021 & \textbf{ZIPPER} & \cite{zhang2021zipper} & & & \centering \checkmark & Tile- and operator-level parallelism \\ \hline
    
\end{tabular}
\end{table}

GraphTheta \cite{li2021graphtheta} also deploys a form of hybrid parallelism to overcome the scalability issue when handling graphs with highly skewed vertex degree distribution. When processing a full iteration with a high-degree vertex, a worker could run out of memory. For that reason, GraphTheta not only splits up and distributes the input graph among workers, but also the operations forming forward and backward pass. This ensures an efficient training phase with natural graphs. 

Data and model parallelism are exploited by ZIPPER \cite{zhang2021zipper}. Subgraphs are formed by applying grid-based partitioning on the adjacency matrix. The partitions are processed in parallel resulting in data parallelism. 
Further, model parallelism is achieved by separating and overlapping the operations forming forward and backward pass. Now, the different operations can be executed concurrently on selected partitions to speed up computation and use the memory more efficiently. Table \ref{tab:parallel} summarizes the types of parallelism.

\subsubsection{Message Propagation} \label{subsubsec:message} 
Similar to the exchange of data in graph processing system, machines in GNN systems may synchronize by \textit{pushing} or \textit{pulling} the relevant values. 
NeuGraph \cite{ma2019neugraph} and GReTA \cite{kiningham2020greta} propagate messages by \textit{pushing} them directly to adjacent vertices. Messages containing graph features and other required information are sent across the network. 
The \textit{pull}-based approach is utilized by systems like DGL \cite{wang2019deep}, Dorylus \cite{thorpe2021dorylus} and GNNAutoScale \cite{fey2021gnnautoscale}. The associated features as well as the k-hop neighborhood are \textit{pulled} from memory to construct the computation graph and perform a training step. However, the moving of features across the network may lead to high communication. 
For that reason, no features are transferred across the network by the P\textsuperscript{3} system \cite{gandhi2021p3}, except for partial activations and error gradients. Moreover, the system proposes a \textit{push-pull} parallelism which switches between \textit{pushing} and \textit{pulling} during the training phase. First, P\textsuperscript{3} \textit{pulls} the desired neighborhood of a vertex to build a computation graph which is \textit{pushed} to all the machines to start the training phase. After having computed the partial activations at Layer 1, the machines \textit{pull} them from all other machines. Then, the computation of the forward pass is performed until the last layer and the backward pass starts.  At Layer 1, the error gradients are \textit{pushed} back to all machines and the backward pass ends. The authors chose to switch between the \textit{push} and \textit{pull} method to decrease the messages transferred over the network.

\subsubsection{Synchronization Mode} \label{subsubsec:sync} 
When performing tasks in parallel, it is important to determine the execution mode. It needs to be decided whether to follow a synchronous, asynchronous or hybrid scheme. As seen in distributed graph processing and distributed NN training, it depends on the specific architecture which mode is best. 
Therefore, AliGraph \cite{yang2019aligraph} does not enforce a particular synchronization method, but chooses the updating mode based on the provided training algorithm. If the implemented algorithm uses synchronous updates, the system will adopt the synchronous scheme, whereas the asynchronous mode will be chosen if the given training algorithm is based on asynchronous execution. Besides AliGraph, GraphTheta \cite{li2021graphtheta} is also not fixed to a specific mode. 

In DistGNN \cite{md2021distgnn}, three update algorithms with varying communication intensity during the aggregation phase are implemented and compared. The update algorithms regard target vertices and their replicas that emerged during vertex-cut partitioning. The first algorithm does not allow any communication between split-vertices in local partitions and their cloned vertices. Hence, there is no need for synchronization. The second algorithm supports communication of local partitions with their replicas, the vertices send partial aggregates to their replicas. Only if all vertices have finished communication, the vertices move to the next step of the training phase. A delayed update mechanism is exploited in the third algorithm. It is an asynchronous execution mode where the vertices send partial aggregates in the current epoch and receive them in a consecutive one. In this way, remote communication and local computation are overlapped. Communication is further avoided by only regarding selected split-vertices during each epoch. The overall results show that the zero-communication strategy is the fastest while maintaining only slight fluctuations in accuracy, followed by the asynchronous delayed update algorithm. 

Dorylus \cite{thorpe2021dorylus} compares three execution mode variants, a synchronous version and two asynchronous ones which differ in the choice of the staleness threshold. Synchronization is performed at the \texttt{gather} operation, meaning if the neighbors of a vertex have not finished to scatter their updated values, the vertex cannot start computing the next layer. For the asynchronous versions, the staleness threshold $s$ determines which stale values of neighbors are allowed to be used by a vertex. For one experiment, the authors chose a value of $s=0$. In this case, stale values of neighbors might be used if the neighbor is in the same epoch. A staleness value of $s=1$ was chosen for another experiment allowing for two successive epochs. By employing asynchronous updates with $s=0$, the per-epoch time could be sped up by around $1.234 \times$. A staleness value that is too high induces slow convergence. Even though the time needed for one epoch decreases with $s=0$ compared to synchronous execution, the number of epochs to obtain the same accuracy rises. 

The execution with P\textsuperscript{3} \cite{gandhi2021p3} needs to be highly coordinated as all machines switch concurrently from data to model parallelism. Additionally, during the data-parallel phase, global gradient synchronization is performed. Therefore, P\textsuperscript{3} follows a synchronous execution mode. 

\subsubsection{Scheduling} \label{subsubsec:schedule} 
To distribute the workload evenly across all workers, tasks and data need to be assigned in an intelligent way. Therefore, scheduling methods are applied which help to increase workload balance and minimize idle times. In the following, we describe important scheduling techniques. An overview and categorization can be found in Table \ref{tab:scheduling}.

\begin{table}[b]
\tiny
\caption{Categorization of scheduling strategies}
\label{tab:scheduling}
\centering
\begin{tabular}{|l|l|l|c|c|l|}
    \hline
    
    & \multicolumn{2}{c|}{\textbf{System}} & \multicolumn{2}{c|}{\textbf{Static vs. Dynamic}} & \\ \cline{4-5}
    
    \textbf{Year} & \multicolumn{2}{c|}{\textbf{System}} & \centering Static & \centering Dynamic & \textbf{Main Concepts} \\ \hline \hline
    
    2019 & \textbf{NeuGraph} & \cite{ma2019neugraph} & & \centering \checkmark & Selective scheduling \\ \hline 
    
    2020 & \textbf{AGL} & \cite{zhang2020agl} & \centering \checkmark & & Parallel preprocessing and model computation stage \\ \hline 

    2021 & \textbf{GraphTheta} & \cite{li2021graphtheta} & & \centering \checkmark & Work stealing technique \\ \hline 

    2021 & \textbf{FlexGraph} & \cite{wang2021flexgraph} & \centering \checkmark & & Computation cost based \\ \hline 

    2021 & \textbf{P\textsuperscript{3}} & \cite{gandhi2021p3} & \centering \checkmark & & Inspired by PipeDream \cite{narayanan2019pipedream}, based on dependencies within computation graph \\ \hline

    2021 & \textbf{Dorylus} & \cite{thorpe2021dorylus} & & \centering \checkmark & Divide tasks based on data and computation type \\ \hline 

    2021 & \textbf{ZIPPER} & \cite{zhang2021zipper} & \centering \checkmark & & Co-locate operations of different subgraphs \\ \hline 
\end{tabular}
\end{table}

The AGL \cite{zhang2020agl} pipeline divides the training procedure into two main stages: a preprocessing stage where the data is loaded and a model computation stage. Instead of performing the two stages sequentially, AGL schedules the stages in parallel. The time needed for the preprocessing stage is smaller compared to the model computation stage. Thus, the training time almost equals the time needed for model computation after some iterations. 

Selective scheduling in NeuGraph \cite{ma2019neugraph} chooses the most important vertices for computing the edge values based on costs for data copy and transfer. Thus, only a subset of vertex data is transmitted to the GPU and unnecessary vertices are not regarded. Further, the system makes use of pipeline scheduling to find the best execution configuration. To hide the transfer latency, transfer of data chunks between host and device memory and computation is overlapped. An initial scheduling plan is iteratively refined during the training process. The initial random order is gradually adjusted by swapping pairs of chunks while monitoring computation and transfer time to ensure an optimal final schedule. 
GraphTheta \cite{li2021graphtheta} adopts a work-stealing scheduling technique \cite{blumofe1999scheduling}. Tasks are assigned to all machines which then start computing. As soon as a machine has finished its tasks, it "steals" tasks that are queued for other machines and processes them. Benefits of this method are improved load balance and efficiency due to reduced idle times.

FlexGraph \cite{wang2021flexgraph} deploys workload balancing using a cost function to reduce communication. In place of metrics like vertex weight or edge weight, the proposed cost function is based upon the GNN training cost per partition. To predict the computation cost of a vertex, features like the number of neighbors as well as the size of each neighborhood are taken into account. The predicted costs of all vertices are summed to estimate the final computation cost of the partition. An online workload balancing strategy uses the estimations to construct a fixed number of balancing plans where certain vertices should be moved from overloaded to other partitions. Finally, the system chooses the plan cutting the least number of edges. For an even more efficient computational process, FlexGraph uses a pipeline processing strategy overlapping computation and communication.

Inspired by PipeDream \cite{narayanan2019pipedream}, P\textsuperscript{3} exploits a simple pipelining mechanism. As soon as a computation phase of a mini-batch is dependent on another phase, communication starts. This communication is overlapped with the computation of other mini-batches to avoid stalls. Due to the pipeline delay, weight staleness occurs. Consequently, a weight update function regarding weights from the previous forward and backward pass is applied. 
Dorylus \cite{thorpe2021dorylus} decomposes forward and backward pass into fine-grained tasks. The tasks are categorized based on \textit{data type} and \textit{computation type}. Depending on the type, the tasks are processed differently and can be performed concurrently. Hereby, communication latency is avoided. Furthermore, the tasks are pooled and whenever a worker is ready, it takes the one that is scheduled next and executes it. 
ZIPPER \cite{zhang2021zipper} co-locates operations of different subgraphs with a pipelining strategy. As different operations target different resources, the overall performance increases due to more efficiently utilized resources.

\subsubsection{Coordination} \label{subsubsec:coord}
Section \ref{back:coord} showed that it is possible to perform the training phase in a centralized, decentralized or hybrid manner. DistDGL \cite{zheng2020distdgl} leaves the choice whether to operate central or decentral to the underlying framework. For example, if DistDGL is built on top of PyTorch \cite{paszke2019pytorch}, an \texttt{all-reduce} primitive is executed to collect and distribute information. However, if the backend framework is TensorFlow \cite{abadi2016tensorflow}, DistDGL supports a parameter server implementation. 

AGL \cite{zhang2020agl} operates central and makes use of a parameter server as introduced in Section \ref{back:coord}. The parameter server stores all required data and features. Each machine accesses it to fetch the assigned graph partition and exchanges updates without extra communication from machine to machine. GraphTheta \cite{li2021graphtheta} also supports computation in a central fashion. Whereas systems like AGL store current model parameters, the parameter server in GraphTheta keeps multiple version of parameters. In this manner, machines can fetch the required parameter version at any time helping to concurrently execute tasks with the appropriate parameters. 
DistGNN \cite{md2021distgnn} shares updates in a decentralized way with an \texttt{all-reduce} operation and direct communication from machine to machine. Consequently, the need for a parameter server is eliminated. Another decentral system is PaGraph \cite{lin2020pagraph}. Here, trainer processes directly interact to exchange locally computed gradients.

\subsubsection{Datasets and Benchmarks}
\begin{table}[t]
    \tiny
    \caption{Overview of graph datasets}
    \label{tab:data}
    \centering
    \begin{tabular}{|l|l|l|l|c|c|c|p{0.35\columnwidth}|}%
    \hline
        
        &&&& \multicolumn{3}{c|}{\textbf{Task Type}} & \\ \cline{5-7}
        
        \textbf{Name} & & \textbf{\#Vertices} & \textbf{\#Edges} & \centering Vertex & \centering Edge & \centering Graph & \textbf{Systems} \\ \hline \hline
        
        \textbf{CiteSeer} & \cite{giles1998citeseer} & 3,327 & 4,732 & \centering \checkmark & \centering \checkmark & & PyG, GraphTheta, GNNAdvisor, GNNAutoScale \\ \hline
        
        \textbf{CORA} & \cite{mccallum2000automating}& 2,708 & 5,429 & \centering \checkmark & \centering \checkmark & & PyG, AGL, GraphTheta, GNNAdvisor, GNNAutoScale \\ \hline 
        
        \textbf{PubMed} & \cite{sen2008collective} & 19,717 & 44,338 & \centering \checkmark & \centering \checkmark & & PyG, NeuGraph, ROC, GraphTheta, GNNAdvisor, GNNAutoScale \\ \hline 
                
        \textbf{PPI} & \cite{zitnik2017predicting, hamilton2017inductive} & 2,373 & 61,318 & \centering \checkmark & & & PyG, ROC, AGL, GNNAdvisor, GNNAutoScale \\ \hline 
        
        \textbf{Reddit} & \cite{hamilton2017inductive} & 232,965 & 114,848,857 & \centering \checkmark & & \centering \checkmark & DGL, GReTA, NeuGraph, ROC, PaGraph, GraphTheta, FlexGraph, Dorylus, DistGNN, DeepGalois \\ \hline
        
        \textbf{LiveJournal} & \cite{yang2015defining} & 4,847,571 & 68,993,773 & & \centering \checkmark & & GReTA, PaGraph, ZIPPER \\ \hline 
        
        \textbf{OGBL-ppa} & \cite{hu2020open} & 576,289 & 30,326,273 & & \centering \checkmark & & DGL \\ \hline
        
         \textbf{OGBL-citation2} & \cite{hu2020open} & 2,927,963 & 30,561,187 & & \centering \checkmark & & DGL \\ \hline
        
         \textbf{OGBN-arxiv} & \cite{hu2020open} & 169,343 & 1,166,243 & \centering \checkmark & & & DGL, GNNAutoScale \\ \hline
        
         \textbf{OGBN-proteins} & \cite{hu2020open} & 132,534 & 39,561,252 & \centering \checkmark & & & DGL, GNNAdvisor \\ \hline
        
         \textbf{OGBN-products} & \cite{hu2020open} & 2,449,029 & 61,859,140 & \centering \checkmark & & & DGL, DistDGL, P3, GNNAutoScale, DistGNN, DeepGalois\\ \hline
        
         \textbf{OGBN-papers100M} & \cite{hu2020open} & 111,059,956 & 1,615,685,872 & \centering \checkmark & & & DistDGL, P3, DistGNN \\ \hline

        \textbf{MAG240M} & \cite{hu2021ogb} & 244,160,499 & 1,728,364,232 & \centering \checkmark & & & - \\ \hline
        
         \textbf{WikiKG90Mv2} & \cite{hu2021ogb} & 91,230,610 & 601,062,811 & & \centering \checkmark & & - \\ \hline
        
         \textbf{PCQM4Mv2} & \cite{hu2021ogb} & 52,970,652 & 54,546,813 & & & \centering \checkmark &  - \\ \hline
         
    \end{tabular}
\end{table}

This section gives an overview of commonly used datasets in the literature to provide a summary of applications and use cases of GNN systems. We highlight selected publicly available graph datasets and show their characteristics (see Table \ref{tab:data}). There are some early citation graph datasets also used to evaluate graph processing systems called CiteSeer \cite{giles1998citeseer}, CORA \cite{mccallum2000automating} and PubMed \cite{sen2008collective}. Here, the vertices represent documents and the edges represent the citations between them. The size is rather small with approximately 3,000 vertices and 5,000 edges in CiteSeer and CORA and  around 19,700 vertices and 44,300 edges in PubMed. Predictions about the vertices and edges can be made, however, no graph-level tasks are currently included. Those sets are included in systems like PyG \cite{fey2019fast}, GraphTheta \cite{li2021graphtheta}, GNNAdvisor \cite{wang2021gnnadvisor} and GNNAutoScale \cite{fey2021gnnautoscale}. 
The Protein-Protein-Interaction (PPI) dataset \cite{zitnik2017predicting, hamilton2017inductive} models the role of proteins in different types of human tissue. It contains 20 graphs, each with an average number of 2,373 vertices and is applicable to vertex-level tasks and is included in systems like PyG \cite{fey2019fast}, ROC \cite{jia2020improving} and AGL \cite{zhang2020agl}.

Nowadays, a common strategy to acquire graph structured data is to crawl social networks and use community information as basis for the resulting graph. About 233,000 Reddit posts from different communities are included in the Reddit \cite{hamilton2017inductive} dataset and the LiveJournal \cite{yang2015defining} set represents around 4.8 million users and their connections. Especially the Reddit dataset is used by numerous systems \cite{wang2019deep, kiningham2020greta, ma2019neugraph, jia2020improving, lin2020pagraph, li2021graphtheta, md2021distgnn} to measure performance. In contrast to the PPI set, vertex- and graph-level tasks may be carried out on the Reddit set.

The Open Graph Benchmark (OGB) \cite{hu2020open} comprises a collection of datasets of varying size, origin and task types. It is differentiated between small, medium and large sets. The small ones consist of up to 170,000 vertices (OGBN-arxiv), the medium of up to 2.9 million vertices (OGBL-citation2) and the large ones of up to 111 million vertices (OGBN-papers100M). Recently, even larger sets have been added in the course of a large-scale challenge \cite{hu2021ogb} to represent real world data. The largest set, namely MAG240M, includes an academic graph representing papers, paper subjects, authors and institutions. To store the approximately 244 million vertices and 1 billion edges, more than 200 GB are needed. The WikiKG90Mv2 knowledge graph consists of 91 million vertices and around 601 million edges resulting in a file size of up to 160 GB. The third set, PCQM4Mv2, is around 8 GB large and contains over 3,700 graphs with a total vertex number of around 53 million and 54 million edges. OGB also provides a unified evaluation and benchmarking suite. In this manner, researchers are able to run, test and compare the performance of their model to the existing state-of-the-art. 

To date, there is no established standard dataset used by all systems for evaluation. Thus, comparison of their performance is difficult. One could categorize the datasets depending on the sizes and task types as done for the OGB datasets to compare the systems. However, the characteristics of the graphs might differ. For example, some graphs with a similar number of vertices may have varying numbers of edges resulting in more sparse or more dense graphs. A special case are fully connected graphs, where each vertex is connected to each other vertex. Another issue is the continuously growing size of real world graphs. If a graph might be suitable for representing real world data right now, it could not be suitable anymore in a couple of years and a new or updated dataset is needed, making comparison of system performance extremely difficult. 

\begin{table}[t]
\tiny
\caption{Overview of availability and compatibility of publicly available systems}
\label{tab:avcompat}
\centering
\begin{tabular}{|l|l|c|c|c|c|c|ll|}
    \hline
    
    & \multicolumn{2}{c|}{} & \multicolumn{2}{c|}{\centering \textbf{Programming language}} & \multicolumn{2}{c|}{\textbf{Hardware}} & & \\ \cline{4-5} \cline{6-7}
    
    \textbf{Year} & \multicolumn{2}{c|}{\textbf{System}} & \centering Python & \centering C/C++ & \centering CPU & \centering GPU & \multicolumn{2}{c|}{\centering \textbf{Compatibility}} \\ \hline \hline
    
    2019 & \textbf{DGL} & \cite{wang2019deep} & \centering \checkmark & \centering \checkmark & \centering \checkmark & \centering \checkmark  & PyTorch, Tensorflow, MXNet & \\ \hline 
    
    2019 & \textbf{PyG} & \cite{fey2019fast} & \centering \checkmark & & \centering \checkmark & \centering \checkmark & PyTorch & \\ \hline  

    2019 & \textbf{AliGraph} & \cite{yang2019aligraph} & \centering \checkmark & \centering \checkmark & \centering \checkmark & \centering \checkmark & PyTorch, Tensorflow & \\ \hline 

    2020 & \textbf{DistDGL} & \cite{zheng2020distdgl} & \centering \checkmark & & \centering \checkmark & & DGL & \\ \hline 

    2021 & \textbf{Dorylus} & \cite{thorpe2021dorylus} & \centering \checkmark & \centering \checkmark & \multicolumn{2}{c|}{serverless} & - & \\ \hline 

    2021 & \textbf{GNNAdvisor} & \cite{wang2021gnnadvisor} & \centering \checkmark & \centering \checkmark & \centering \checkmark & \centering \checkmark & PyG, DGL, Gunrock & \\ \hline 
    
    2021 & \textbf{GNNAutoScale} & \cite{fey2021gnnautoscale} & \centering \checkmark & \centering \checkmark & \centering \checkmark & \centering \checkmark & PyG & \\ \hline 

\end{tabular}
\end{table}

\subsubsection{Availability and Compatibility}
To better asses which system to choose, it is important to know which ones are freely accessible and which frameworks are compatible (see Table \ref{tab:avcompat}). We found seven systems to be publicly available, namely PyG \cite{fey2019fast}, DGL \cite{wang2019deep}, AliGraph \cite{yang2019aligraph}, DistDGL \cite{zheng2020distdgl}, GNNAdvisor \cite{wang2021gnnadvisor}, GNNAutoScale \cite{fey2021gnnautoscale} and Dorylus \cite{thorpe2021dorylus}. PyG is a library built upon PyTorch \cite{paszke2019pytorch} and is fully implemented in Python. It comes with multi-GPU support to achieve scalability. Also, there is a GitHub community\footnote{\url{https://github.com/pyg-team/pytorch_geometric}} with over 240 contributors. In contrast to PyG, DGL is framework agnostic and GNN models can be built with PyTorch, Tensorflow \cite{abadi2016tensorflow} or Apache MXNet \cite{chen2015mxnet}. DGL includes CPU and GPU support and is implemented in Python and C++. The GitHub community\footnote{\url{https://github.com/dmlc/dgl}} includes almost 200 contributors. DistDGL is integrated in DGL as a module\footnote{\url{https://docs.dgl.ai/en/0.6.x/api/python/dgl.distributed.html}}. The used programming language is Python and it runs on a cluster of CPUs. AliGraph\footnote{\url{https://github.com/alibaba/graph-learn}} is compatible with PyTorch and Tensorflow and uses Python and C++. GNNAdvisor\footnote{\url{https://github.com/YukeWang96/OSDI21_AE}} is also implemented with Python and C++. It can either be built upon DGL, PyG or Gunrock \cite{wang2016gunrock} as underlying framework and includes CPU and GPU support. GNNAutoScale, also referred to as PyGAS, is implemented in PyTorch and uses PyG. The code can be downloaded on GitHub\footnote{\url{https://github.com/rusty1s/pyg_autoscale}}. Dorylus combines data servers with serverless computing. The main logic is written in Python and C++, the code is available on GitHub\footnote{\url{https://github.com/uclasystem/dorylus}}. 
Besides the target use case, it is important to be aware of the framework support as well as utilized programming languages when deciding on a system. Some systems support various underlying frameworks, for instance DGL, AliGraph or GNNAdvisor. Other systems are bound to a specific set up, for example GNNAutoScale is built upon PyG, and PyG is implemented on top of PyTorch. Ultimately, the user needs to decide which one suits the application and the personal preferences the best.

\subsubsection{Performance assessment}
After reviewing the optimizations, availabilities and compatibilities of a plethora of systems, it would be interesting to understand the performance behavior of each approach.  
As a full fledged benchmarking effort is beyond the scope of this survey, we now present a selection of performance results available from the literature. 
DistDGL \cite{zheng2020distdgl} gains an overall speedup of $2.2\times$ over Euler \cite{alibaba2020euler} and is more than $5\times$ faster in the data copy phase. GNNAdvisor \cite{wang2021gnnadvisor} outperforms DGL \cite{wang2019deep} by a factor of $3$ and NeuGraph \cite{ma2019neugraph} by a factor of up to $4$ when measuring training time. When evaluating ROC \cite{jia2020improving}, the system achieves to perform up to $4\times$ higher as measured in number of epochs per second in the given experiments than NeuGraph. ROC, in turn, is outperformed by P\textsuperscript{3} \cite{gandhi2021p3} which completes epochs up to $2\times$ faster. 
Although some general points can be made above about how well the systems perform, the used datasets and hardware setups differ across the reported systems. This makes it difficult to draw explicit conclusions from the reported quantifications across different papers. Consequently, the need for a comprehensive performance comparison of GNN systems arises, a worthy endeavor for future work.

\begin{table}[b]
    \tiny
    {
    \caption{Connections across systems for graph processing, DL and GNNs}
    \label{tab:finaloverview}
    \centering
    \resizebox{0.7\columnwidth}{!}{
    \begin{tabular}{|c|c|c|c|}
        \hline
         & \textbf{Graph Processing Systems} & \textbf{DL Systems} & \textbf{GNN Systems} \\ \hline \hline
         \textbf{Partitioning} & Section \ref{back:part} & - & Section \ref{subsubsec:part} \\ \hline
         \textbf{Sampling} & Section \ref{back:sampling} & - & Section \ref{subsubsec:sampling} \\ \hline
         \textbf{Programming Abstraction} & Section \ref{back:distgp} & - & Section \ref{subsubsec:api} \\ \hline
         \textbf{Inter-Process Communication} & Section \ref{back:inter} & - & Section \ref{subsubsec:inter} \\ \hline
         \textbf{Parallelism} & - & Section \ref{back:par} & Section \ref{subsubsec:parallel} \\ \hline
         \textbf{Message Propagation} & Section \ref{back:message} & - & Section \ref{subsubsec:message} \\ \hline
         \textbf{Synchronization Mode} & Section \ref{back:syncgp} & Section \ref{back:syncdl} & Section \ref{subsubsec:sync} \\ \hline
         \textbf{Scheduling} & Section \ref{back:syncgp} & - & Section \ref{subsubsec:schedule} \\ \hline
         \textbf{Coordination} & - & Section \ref{back:coord} & Section \ref{subsubsec:coord} \\ \hline
    \end{tabular}
    }}
\end{table}

\subsection{Connections across graph processing, deep learning and GNN systems}
We have introduced methods used to distribute and optimize graph processing (Section \ref{back:distgp}) and DL (Section \ref{back:nntraining}) and techniques commonly used by systems for GNNs in Section \ref{gnn:methods}. Table \ref{tab:finaloverview} shows how the introduced methods and the systems are connected. Systems for GNNs and systems for graph processing share a wide range of principles, such as partitioning, message propagation and scheduling. DL systems share ideas of parallelism, synchronization and coordination with GNN systems. Sampling occurs in GNN training and in graph processing. However, sampling in GNN training is not exactly the same as sampling in graph processing. In GNN training, sampling is used to exclude certain vertices during a training epoch and to create mini-batches, while sampling in graph processing means to sparsify the whole input graph before starting the computation.

\section{Discussion and Outlook} \label{subsec:discussion} 
We have seen that designing systems for GNN training is a challenging task. The recently proposed systems face issues like workload imbalance, redundant vertex accesses, communication overhead or changing the parallelism. In the following, we discuss current topics that are increasingly investigated as well as open challenges that still arise when developing GNN systems. 

\subsection{Current Research Trends}
An increasing amount of research dealing with the co-design of software and hardware to accelerate GNN training \cite{zhang2020hardware, chen2021rubik, kiningham2020grip, kiningham2020greta}. Here, not only software and algorithms are optimized, but also hardware modules are developed to better address the characteristics of GNNs. 
Another interesting approach examines the acceleration of quantized GNN architectures \cite{wang2022qgtc}. Quantized GNNs \cite{tailor2020degree, feng2020sgquant, ding2021vq, saad2021quantization} have increasingly emerged in the past years and incorporate compressed weights and node embeddings to reduce the memory footprint and computation. In addition, they are robust and the loss of accuracy is marginal. Systems adapted to quantized GNNs can further accelerate training and inference times. 
Beyond quantized GNNs, acceleration techniques for dynamic graphs have been proposed \cite{guan2022dynagraph, wang2021apan}. Dynamic graphs change over time, edges might be added or deleted and vertex features evolve. Some architectures exist that are able to process dynamic graphs, for example, SDG \cite{fu2021sdg}, Dynamic-GRCNN \cite{peng2020spatial}, and DyGNN \cite{ma2020streaming}. Dynamic graphs and architectures for processing them pose new research challenges that are currently being solved. 
Ongoing research also aims at optimizing inference for GNNs \cite{zhang2022graphless, gao2022efficient}, as real-time inference might be crucial for applications such as self-driving cars \cite{zhou2021accelerating}. Methods range from combining multi-layer perceptrons \cite{jain1996artificial} with GNNs \cite{zhang2022graphless}, introducing a novel propagation scheme which is adapted per node \cite{gao2022efficient} and pruning channels of the feature matrices which leads to pruned weight matrices \cite{zhou2021accelerating}. 
Moreover, work focusing on improving the data loading and preprocessing step is proposed \cite{liu2021bgl, jangda2021accelerating}. When measuring the time of the different training steps, it can be noted that a prominent proportion of the overall training time is consumed by reading in and preprocessing the graphs. By investigating and improving data I/O and preprocessing, the training process can be made more efficient in future work. 
Lately, neural architecture search (NAS) and automated methods for GNNs has caught the attention of many experts \cite{gao2020graph, cai2021rethinking, li2020autograph, huan2021search, zhao2020learned}. Going a step further, mechanisms like NAS can inspire researchers to also include automated methods assessing which optimization technique fits best for the given architecture, data and system. For instance, it could be automatically made a recommendation which partitioning strategy is most suitable depending on the requirements of the user. But not only single analyses could be made, also an overall estimation about the combination and composition of optimizations. 

\subsection{Open Challenges} 
As GNNs and systems for GNNs are young research fields, there still are open challenges to be solved. Most systems are evaluated on static graphs and standard GNN models like GCN, GAT and GraphSAGE, but there is no information on how the systems would perform on more specialized models and types of data. There are, for instance, GNNs for hypergraphs \cite{feng2019hypergraph, yadati2019hypergcn} and multigraphs \cite{ouyang2019learning, geng2019spatiotemporal}. As they come with different characteristics than general GNNs, it is likely that optimization techniques need to be adapted to those types of GNNs. 
Beyond that, deeper GNN architectures with numerous layers have recently been proposed \cite{li2020deepergcn, li2021training, liu2020towards} which could pose new challenges for developing GNN systems. 
Another topic is the size of the feature vectors. When evaluating the systems, fixed feature vector lengths are used, but there is no consistent size within the presented systems. Now, the question is whether the systems perform equally well with variable feature vector sizes as the size of the feature vector might have a great effect on storage, data loading and computation time \cite{huang2021understanding}. 
Moreover, the presented systems are all either based upon an existing graph processing system or a DL framework. Then, the missing operations needed for GNN training are added and optimized. Although this is an effective approach to design a GNN system, the characteristics of GNNs are not fully exploited. A GNN training iteration alternately incorporates dense computation of NN operations with regular memory access and a sparse aggregation phase with irregular memory accesses \cite{wang2021gnnadvisor}. Furthermore, the first layers of a GNN generally are the most compute intensive ones. Hence, it could be beneficial to adapt the system to handle the computations of the first layers differently compared to the subsequent ones. P\textsuperscript{3} is the first system going this direction, but it is difficult to find the optimal spot when to change from model to data parallelism and vice versa. In addition, finding a good way to partition and distribute the model among the machines is challenging. In general, such a method is not favorable if the underlying assumption that the activations are significantly smaller than the features is not met.  

\section{Conclusions}

GNNs are increasingly used in various fields of application. With the steadily growing size of real-world graphs, the need for large-scale GNN systems arises. To better comprehend the main concepts of GNN systems, we first provided an overview of two fundamental topics, systems for graph processing and systems for DNN training. We established connections between the used methods and showed that many ideas of GNN systems are inspired by the two related fields. In the main part of this survey, we discussed, categorized and compared concepts for distributed GNN training. This included partitioning strategies, sampling methods, different types of parallelism as well as efficient scheduling and caching. We further investigated datasets and benchmarks for evaluation as well as availability and compatibility of the systems. 
Although the current systems are able to scale to large graphs, there still are unresolved issues. For instance, the support for specialized GNN architectures like dynamic GNNs or those using hierarchical aggregation are not fully explored. Further, it should be investigated how to handle variable feature vector sizes as the systems only provide insights into a fixed length. 
When looking into the future, we see a trend in investigating system support for quantized GNNs as they are robust and only have a small memory footprint. Another interesting type of GNN architecture that could be increasingly supported are dynamic GNNs. The software-hardware co-design of systems is receiving growing attention and systems even more adapted to the characteristics of GNNs can further improve the performance. Lastly, a closer look at the data loading and preprocessing phase can help to minimize the overall training time.


\bibliographystyle{ACM-Reference-Format}
\bibliography{bib}


\end{document}